\documentclass[a4paper,11pt]{article}

\usepackage{jcappub} 
\usepackage{color}
\usepackage{xcolor}
\usepackage[T1]{fontenc} 

\newcommand\sch{Schwarzschild}
\newcommand\schbh{Schwarzschild black hole}

\newcommand\rn{Reissner-Nordstr\"{o}m}
\newcommand\rnbh{Reissner-Nordstr\"{o}m black hole}

\newcommand\hornbh{Horndeski black hole}
\newcommand\galbh{Galileon black hole}

\title{\boldmath Weak and strong deflection gravitational lensings by a charged Horndeski black hole}

\author{Cheng-Yi Wang,}
\author{Yu-Fu Shen}
\author[1]{and Yi Xie\note{Corresponding author.}}

\affiliation{School of Astronomy and Space Science, Nanjing University, Nanjing 210023, China\\
Key Laboratory of Modern Astronomy and Astrophysics, Nanjing University, Ministry of Education, Nanjing 210023, China}

\emailAdd{yixie@nju.edu.cn}

\abstract{A charged black hole was predicted by the Einstein--Horndeski--Maxwell theory. In order to provide its observational signatures, we investigate its weak and strong deflection gravitational lensings. We find its weak deflection lensing observables, including the positions, magnifications and differential time delay of the lensed images. We also obtain its strong deflection lensing observables, including the apparent radius of the photon sphere as well as the angular separation, brightness difference and differential time delay between the relativistic images. Taking the supermassive black hole in the Galactic Center as the lens, we evaluate these observables and compare these signatures with those of the Schwarzschild, Reissner-Nordstr\"{o}m, tidal Reissner-Nordstr\"{o}m and charged Galileon black holes. After a detailed analysis of the feasibility of measuring these lensing observables, we conclude that although it is possible to detect some leading effects of the weak and strong deflection lensings by the charged Horndeski and other black holes with current technology, it would be unlikely to distinguish one kind of these black holes from the others based on these detections in the near future due to lack of enough highly angular resolution in astronomical observations to tell their differences.  
}


\begin{document}
\maketitle
\flushbottom
\allowdisplaybreaks

\section{Introduction}

\label{sec:intro}

As the most general scalar-tensor theory of gravitation in four dimensional spacetime yielding second-order field equations, the Horndeski theory was constructed in 1974 \cite{Horndeski1974IJTP10.363} by inspiration of the work of Lovelock \cite{Lovelock1971JMP12.498}. It is recently revived again in the aftermath of severe attack from gravitational wave detection \cite{Rham2018PRL121.221101}. The discovery of cosmic acceleration \cite{Riess1998AJ116.1009,Perlmutter1999ApJ517.565} has triggered an extensive search of modifications of the general relativity (GR) \cite{Clifton2012PhR513.1}, in which the scalar-tensor theories are regarded as the most natural, simplest and consistent modification of GR and the most well studied and established alternative theories. In the last decade, the Horndeski theory has played a major role as a ghost-free effective field theory acting as dark energy because it can include GR, Brans-Dicke theory \cite{Brans1961PR124.925}, the $f(R)$ gravity \cite{Felice2010LRR13.3,Sotiriou2010RMP82.451}, the Dvali--Gabadadze--Porrati model \cite{Dvali2000PLB485.208} and the Galileon model \cite{Deffayet2009PRD79.084003,Deffayet2009PRD80.064015,Kobayashi2011PTP126.511,Deffayet2013CQG30.214006} in some specific limits. However, after the direct detections of gravitational waves and optical counterpart of a binary neutron star merger \cite{Abbott2017PRL119.161101,Abbott2017ApJ848.L12,Abbott2017ApJ848.L13}, the extremely small difference between the phase velocity of the gravitational wave and the speed of light almost ruled out the whole Horndeski theory as a candidate of dark energy \cite{Lombriser2017PLB765.382,Lombriser2016JCAP03.031,Baker2017PRL119.251301,Creminelli2017PRL119.251302,Sakstein2017PRL119.251303,Ezquiaga2017PRL119.251304} but with few exceptions \cite{Casalino2018PDU22.108,Casalino2019CQG36.017001} (see ref. \cite{Ishak2019LRR22.1} for a recent review). Nevertheless, it was currently found \cite{Rham2018PRL121.221101} that the energy scale observed at LIGO is very close to the cutoff of the Horndeski theory and the operators at the cutoff scale can bring the speed of gravitational waves to the speed of light at LIGO scale, making this theory alive again.

Astrophysics under the Horndeski theory also drew much attention. Its effects on the motions of planets and pulsars and light propagation were constrained in the Solar and stellar systems \cite{Hohmann2015PRD92.064019,Bhattacharya2017PRD95.044037,Hou2018EPJC78.247,Dyadina2019MNRAS483.947}. While stars were obtained in this theory \cite{Cisterna2015PRD92.044050,Cisterna2016PRD93.084046,Babichev2016CQG33.154002,Maselli2016PRD93.124056,Silva2016IJMPD25.1641006} and proved to be absence of scalar hair \cite{Lehebel2017JCAP07.037}, Horndeski black holes \cite{Rinaldi2012PRD86.084048,Anabalon2014PRD89.084050,Cisterna2014PRD89.084038,Babichev2015JCAP5.31,Babichev2016JCAP09.011,Babichev2017JCAP04.027,Antoniou2018PRD97.084037,Bakopoulos2019PRD99.064003} are more attractive since they might be able to evade the no-hair theorems \cite{Hui2013PRL110.241104,Sotiriou2014PRL112.251102,Sotiriou2014PRD90.124063,Antoniou2018PRL120.131102}. Stabilities \cite{Anabalon2014PRD90.124055,Cisterna2015PRD92.104018} and thermodynamics \cite{Miao2016EPJC76.638,Feng2016PRD93.044030} of these black holes were also investigated. 

As a powerful tool in gravitational physics \cite{Perlick2004LRR7.9}, signals of both weak and strong deflection gravitational lensing carry an abundance of information on fundamental physics of black holes and nature of gravity, which might be taken to test the Horndeski theory in a different way from the gravitational waves. Weak deflection lensing has been employed to provide insights on astronomy, astrophysics and cosmology \cite{Schneider1992GL,Petters2001STGL,Schneider2006GLSWM,Aubourg1993Nature365.623,Sahu2017Sci356.1046}, to test GR and the modified theories of gravity \cite{Keeton2005PRD72.104006,Keeton2006PRD73.044024,Keeton2006PRD73.104032,Collett2018Sci360.1342} and to probe the interaction between electromagnetic and gravitational fields \cite{Li2017AoP382.136,Cao2018EPJC78.191}. Strong deflection lensing can generate a ``shadow'' and relativistic images caused by photons winding several loops around a black hole, which is a unique feature in the nearby region around the black hole and is not presented in the weak deflection lensing  (see \cite{Bozza2010GRG42.2269,Eiroa2012arXiv1212.4535,Cunha2018GRG50.42} for reviews). The photon sphere of a black hole plays an important role in the strong deflection. The Event Horizon Telescope \footnote{\url{https://eventhorizontelescope.org/}} (EHT) is directly observing and imaging the shadows of the supermassive black hole Sgr A* in the Galactic Center as well as the supermassive black hole at the center of M87. Strong deflection lensings by static and symmetric black holes \cite{Darwin1959PRSLSA249.180,Luminet1979AA75.228,Ohanian1987AJP55.428,Nemiroff1993AJP61.619,Bozza2001GRG33.1535,Virbhadra1998AA337.1,Virbhadra2000PRD62.084003,Bozza2002PRD66.103001,Eiroa2002PRD66.024010,Bhadra2003PRD67.103009,Bozza2004GRG36.435,Perlick2004PRD69.064017,Whisker2005PRD71.064004,Majumdar2005IJMPD14.1095,Eiroa2004PRD69.063004,Eiroa2005PRD71.083010,Eiroa2006PRD73.043002,Iyer2007GRG39.1563,Mukherjee2007GRG39.583,Bozza2008PRD78.103005,Virbhadra2008PRD77.124014,Virbhadra2009PRD79.083004,Bin-Nun2010PRD82.064009,Eiroa2012PRD86.083009,Eiroa2013PRD88.103007,Zhao2017EPJC77.272,Zhao2017PLB774.357} and rotating black holes \cite{Bozza2003PRD67.103006,Vazquez2004NCimB119.489,Bozza2005PRD72.083003,Bozza2006PRD74.063001,Bozza2007PRD76.083008,Bozza2008PRD78.063014,Chen2010CQG27.225006,Chen2011PRD83.124019,Kraniotis2011CQG28.085021,Chen2012PRD85.124029,Kraniotis2014GRG46.1818,Ji2014JHEP03.089,Cunha2015PRL115.211102,Wang2016JCAP11.020} are widely studied and they can be used to discriminate black holes \cite{Bin-Nun2010PRD81.123011, Gyulchev2007PRD75.023006, Gyulchev2013PRD87.063005, Zhao2016JCAP07.007, Cavalcanti2016CQG33.215007,Chakraborty2017JCAP07.045}, naked singularities \cite{Gyulchev2008PRD78.083004,Sahu2012PRD86.063010,Sahu2013PRD88.103002,Virbhadra2002PRD65.103004} and wormholes \cite{Kuhfittig2014EPJC74.2818, Kuhfittig2015arXiv1501.06085,Nandi2006PRD74.024020,Tsukamoto2012PRD86.104062,Tsukamoto2016PRD94.124001} as well as test gravity \cite{Eiroa2014EPJC74.3171,Sotani2015PRD92.044052} and its fundamental interaction with electromagnetic field \cite{Chen2015JCAP10.002,Lu2016EPJC76.357,Chen2017PRD95.104017,Zhang2018EPJC78.796,Chen2018EPJC78.981}. Strong deflection lensings by a neutral Horndeski black hole \cite{Badia2017EPJC77.779} and by a charged \galbh\ \cite{Zhao2016JCAP07.007} were also studied. Nevertheless, the relativistic images of strong deflection lensing are extremely faint \cite{Virbhadra1998AA337.1,Virbhadra2000PRD62.084003,Petters2003MNRAS338.457} and significantly hard to detect for current stage of technology. Therefore, a combination of weak and strong deflection lensing might be complementary to each other and be able to provide a whole picture of a black hole's lensing signatures \cite{Eiroa2002PRD66.024010,Horvath2011PRD84.083006,Eiroa2012PRD86.083009}. 

In this work, we will focus on the weak and strong deflection gravitational lensings by a charged \hornbh. It is an asymptotically flat solution to a sub-class of the Horndeski theory with the scalar field coupled to the background only through the Einstein tensor in the presence of an electric field \cite{Cisterna2014PRD89.084038}. It is more complicated than the charged \galbh\ considered in Ref. \cite{Zhao2016JCAP07.007} which is a perturbative solution \cite{Babichev2015JCAP5.31}. As we will show, their lensing signatures have some similarity in a manner but with enough difference to tell. It is widely believed that an astrophysical black hole in the real universe must have zero charge because the surrounding plasma can neutralize it. However, the electric charge of a black hole can be inherited from its charged collapsed progenitor \cite{Ray2003PRD68.084004}, be acquired by accretion of charged matter and be induced by its rotation in the external magnetic field \cite{Wald1974PRD10.1680}. It was found \cite{Zajacek2018MNRAS480.4408} that Sgr A* may have a small, transient positive charge.

In Sect. \ref{sec:metriclb}, the spacetime of the charged \hornbh\ is briefly reviewed and the set-up for gravitational lensing is given, including its lens equation as well as its exact forms of the bending angle and travelling time span for a photon.  In Sect. \ref{sec:wdl}, we obtain its weak deflection lensing observables, including positions, magnifications and differential time delay of the lensed images. In Sect. \ref{sec:sdl}, the strong deflection lensing observables are found out for the charged \hornbh, including the apparent size of the photon sphere as well as the angular separation, brightness difference and differential time delay between the relativistic images. These results are compared with those of the \sch, \rn, tidal \rn\ and charged \galbh s in both of scenarios of lensings and their feasibility of detection are also discussed in these two sections. Finally, in Sect. \ref{sec:con}, we summarize and discuss our results.

\section{Metric and gravitational lensing set-up}

\label{sec:metriclb}

\subsection{Metric}

In a four dimensional spacetime, the Lagrangian of the Horndeski theory has a form as \cite{Horndeski1974IJTP10.363,Anabalon2014PRD89.084050,Cisterna2014PRD89.084038}
\begin{eqnarray}
  \label{}
  L & = & \lambda_1 \delta^{\alpha\beta\gamma\kappa}_{\rho\sigma\mu\nu}R^{\rho\sigma}_{\alpha\beta}R^{\mu\nu}_{\gamma\kappa} + \lambda_2\delta^{\alpha\beta\gamma}_{\rho\mu\nu}\nabla_{\alpha}\phi\nabla^{\rho}\phi R^{\mu\nu}_{\beta\gamma} + \lambda_3\delta^{\alpha\beta}_{\mu\nu}R^{\mu\nu}_{\alpha\beta} \nonumber\\
  & & + \Theta +B\epsilon^{\alpha\beta\gamma\kappa}R^{\mu}_{\nu\alpha\beta}R^{\nu}_{\mu\gamma\kappa}
\end{eqnarray}
where $R^{\mu}_{\nu\alpha\beta}$ is the Riemann curvature tensor, $B$ is a constant, $\lambda_{1,2,3}$ are arbitrary functions of the scalar field $\phi$ and $\Theta$ is an arbitrary function of $\nabla_{\alpha}\phi\nabla^{\alpha}\phi$ and $\phi$. One particular case of this theory is the scalar field couple to the background only with the Einstein tensor $G_{\mu\nu}$ controlled by parameter $\eta$ in the presence of an electromagnetic field $F^{\mu\nu}$, which reads as ($G=c=1$) \cite{Cisterna2014PRD89.084038} 
\begin{equation}
  \label{eq:action}
  I = \int \sqrt{-g}\mathrm{d}^4x\bigg[ \frac{R}{16\pi } + \frac{\eta}{2}G_{\mu\nu}\nabla^{\mu}\phi\nabla^{\nu}\phi - \frac{1}{4} F_{\mu\nu}F^{\mu\nu}  \bigg].
\end{equation}
It has an asymptotically flat solution describing the charged \hornbh\ as
\begin{equation}
\label{eq:metric}
  \mathrm{d}s^{2} = -A(r)\mathrm{d}t^{2}+B(r)\mathrm{d}r^{2}+C(r)(\mathrm{d}\theta^{2}+\sin^{2}\theta\mathrm{d}\varphi^{2})
\end{equation}
with the functions $A(r)$, $B(r)$ and $C(r)$ as 
\begin{eqnarray}
\label{}
  A(r) & = & 1-\frac{2m_{\bullet}}{r} + \frac{Q}{r^2} - \frac{Q^2}{12r^4} \\
\label{}
  B(r) & = & \bigg(1-\frac{Q}{2r^2}\bigg)^2 [A(r)]^{-1} \\
  \label{}
  C(r) & = &  r^2.
\end{eqnarray}
where $m_{\bullet}$ is the mass and $Q=q_{\mathrm{e}}^2/4$ is an integration constant connected with the charge. In order to ensure the existence of the event horizon and the curvature singularities inside the horizon, the condition that \cite{Feng2016PRD93.044030}
\begin{equation}
  \label{}
  0<Q<\frac{9}{8}m^2_{\bullet}
\end{equation}
has to be satisfied. 

The charged \hornbh\  \eqref{eq:metric} is different from a \rnbh\ \cite{Reissner1916AnP355.106,Nordstrom1918KNAB20.1238}
\begin{equation}
  \label{}
  A_{\mathrm{RN}}(r)=[B_{\mathrm{RN}}(r)]^{-1} = 1-\frac{2m_{\bullet}}{r} + \frac{Q}{r^2},\quad C_{\mathrm{RN}}(r)=r^2,\quad \mathrm{with}\ 0<Q<m^2_{\bullet}
\end{equation}
by an extra term of $r^{-4}$ in $A(r)$ and $A(r)B(r)\neq 1$. It was also found \cite{Feng2016PRD93.044030} that its thermodynamical behavior is different from the one of a \rnbh\ by the fact that its temperature is always positive but not zero. As a solution in the braneworld paradigm \cite{Maartens2004LRR7.7}, the tidal \rnbh\ \cite{Dadhich2000PLB487.1} has the same metric as the \rnbh\ does in the formality. The tidal charge $Q$ controls the $r^{-2}$ terms in both of the metrics. While $Q$ in the \rnbh\ is the square of its electric charge and is always positive, $Q$ in the tidal \rnbh\ can be negative because of the gravitational effects from the fifth dimension \cite{Dadhich2000PLB487.1}.  As a perturbative solution to the theory \eqref{eq:action}, the charged \galbh\ \cite{Babichev2015JCAP5.31} is more similar to the charged Horndeski black hole and has the metric with positive $Q$ as
\begin{eqnarray}
  \label{}
  A_{\mathrm{CG}}(r) & = & 1-\frac{2m_{\bullet}}{r} + \frac{Q}{r^2} \\
  B_{\mathrm{CG}}(r) & = & \bigg(1+ \frac{Q}{r^2}\bigg)^{-1} [A_{\mathrm{CG}}(r)]^{-1} \\
  C_{\mathrm{CG}}(r) & = & r^2,
\end{eqnarray}
with
\begin{equation}
  \label{}
   0<Q<m^2_{\bullet}.
\end{equation}

\subsection{Gravitational lensing set-up}

Although based on quite different contexts, the mathematical descriptions of these four black hole solutions, i.e., the charged Horndeski, \rn, tidal \rn\ and charged \galbh, are close to each other. It will be theoretically and practically interesting to know their observational signatures that might be helpful to distinguish them. In this work, we focus on their gravitational lensings, especially by a supermassive black hole, because ground-based infrared and radio interferometry \cite{GRAVITYCol2017A&A602.A94,Doeleman2008Nature455.78} have continuously improved capabilities to probe lensing signals of Sgr A* and M87 and physical processes nearby them. Detection of gravitational waves from supermassive black hole binaries may have to wait for next generation space-borne detectors.      

For an isolated black hole, the exact bending angle of a deflected light ray can be obtained as  \cite{Virbhadra1998AA337.1,Weinberg1972Book}
\begin{equation}
  \label{eq:dfagl}
  \hat{\alpha}(r_0)= 2\int^{\infty}_{r_0} \frac{\sqrt{B(r)}}{ \sqrt{C(r)} \sqrt{ \frac{C(r)}{C(r_0)} \frac{A(r_0)}{A(r)}-1 } } \mathrm{d}r - \pi,
\end{equation}
where $r_0$ is the closet approach distance of the photon to the black hole. When $r_0\gg2m_{\bullet}$, $\hat{\alpha}$ will be much less than 1 and this integral can be handled in the small angle approximation, giving the weak deflection lensing. As $r_0$ decreases to $2m_{\bullet}$, $\hat{\alpha}$ increases and diverges eventually, giving the strong deflection lensing. In order to find the lensing observables, we take the lens equation as \cite{Virbhadra1998AA337.1,Virbhadra2000PRD62.084003}
\begin{equation}
  \label{eq:lenseq}
  \tan \mathcal{B} = \tan \vartheta - D[\tan \vartheta+\tan(\hat{\alpha}-\vartheta)],
\end{equation}
where $\mathcal{B}$ is the angular position of the source, $\vartheta$ is the angular position of the image, $D=D_{\mathrm{LS}}/D_{\mathrm{OS}}$ and $D_{\mathrm{LS}}$ and $D_{\mathrm{OS}}$ are the angular diameter distances from the lens to the source and from the observer to the source. For a lensed image, its (signed) magnification $\mu$ can be found as \cite{Refsdal1964MNRAS128.295}
\begin{equation}
  \label{}
  \mu(\vartheta) = \bigg[ \frac{\sin\mathcal{B}(\vartheta)}{\sin\vartheta}   \frac{\mathrm{d}\mathcal{B}(\vartheta)}{\mathrm{d}\vartheta}\bigg]^{-1}.
\end{equation}
If the brightness of a source is variable, the differential time delay between its lensed images would be possibly measurable. Its value can be obtained by making use of the total time span for a photon traveling from the source to the observer \cite{Weinberg1972Book,Bozza2004GRG36.435,Virbhadra2008PRD77.124014}
\begin{equation}
  \label{eq:T}
  T =  T(R_{\mathrm{src}}) + T(R_{\mathrm{obs}}),
\end{equation}
with
\begin{equation}
  \label{eq:T(R)}
  T(R)  =  \int^{R}_{r_0}\bigg|\frac{\mathrm{d}t}{\mathrm{d}r}\bigg|\mathrm{d}r
\end{equation}
and
\begin{equation}
  \label{eq:dtdr}
  \frac{\mathrm{d}t}{\mathrm{d}r} = {\sqrt{B(r)C(r)A(r_0)}\over{A(r)\sqrt{C(r_0)}\sqrt{{C(r)\over{C(r_0)}}{A(r_0)\over{A(r)}}-1}}},
\end{equation}
where $R_{\mathrm{obs}}=D_{\mathrm{OL}}$ is the distance from the observer to the lens and $R_{\mathrm{src}}=(D_{\mathrm{OS}}^2\tan^2\mathcal{B}+D_{\mathrm{LS}}^2)^{1/2}$ is the radial coordinate of the source with respect to the lens. 

The bending angle, the lens equation, the magnification and the time delay will be evaluated in both weak and strong deflection lensings and their resulting observables will be discussed for a specific case of Sgr A*.

\section{Weak deflection lensing}

\label{sec:wdl}

In the scenario of weak deflection lensing, the closet approach distance of a light ray to the lens is much larger than its gravitational radius, $r_0\gg m_{\bullet}$, so that no photon in the ray from the source to the observer winds around the lens. Therefore, deflection angle caused by the charged \hornbh\ can be written in the form of a series for weak deflection lensing as 
\begin{equation}
  \label{eq:hatalphab}
  \hat{\alpha}(u) = 4\frac{m_{\bullet}}{u} + \bigg(\frac{15}{4}-q\bigg)\pi \bigg(\frac{m_{\bullet}}{u}\bigg)^2  + \frac{8}{3}(16 - 7 q) \bigg(\frac{m_{\bullet}}{u}\bigg)^3 +\mathcal{O}\bigg(\frac{m_{\bullet}^4}{u^4}\bigg)
\end{equation}
where $u$ is the impact parameter satisfying $C(r_0)=u^2A(r_0)$ and $q\equiv Q/m_{\bullet}^2$. 

For convenience in the following works, we define scaled variables \cite{Keeton2005PRD72.104006,Keeton2006PRD73.044024,Keeton2006PRD73.104032} 
\begin{equation}
  \label{eq:scaledvar}
  \beta = \frac{\mathcal{B}}{\vartheta_{\mathrm{E}}},\qquad \theta=\frac{\vartheta}{\vartheta_{\mathrm{E}}},\qquad \hat{\tau}=\frac{\tau}{\tau_{\mathrm{E}}}, \qquad \varepsilon = \frac{\vartheta_{\bullet}}{\vartheta_{\mathrm{E}}},
\end{equation}
where $\vartheta_{\bullet}=\arctan(m_{\bullet}/D_{\mathrm{OL}})$ is the angular gravitational radius at distance $D_\mathrm{OL}$, $\tau$ is the time delay between images, the angular Einstein ring radius is
\begin{equation}
  \label{}
  \vartheta_{\mathrm{E}} = \sqrt{\frac{4m_{\bullet}D_{\mathrm{LS}}}{D_{\mathrm{OL}}D_{\mathrm{OS}}}}
\end{equation}
and the time scale is
\begin{equation}
  \label{}
  \tau_{\mathrm{E}}=4m_{\bullet}.
\end{equation}
We also assume the observer is far from the lens so that $\varepsilon$ can serve as a small parameter to obtain higher-order observables for the charged \hornbh. These next-to-leading-order observables will be critical for distinguishing various charged black holes by the weak deflection lensing since their leading signals are as the same as those of the \schbh\ in GR (see below for details). 

After a comparison of lensing observables by the charged Horndeski, \rn, the tidal \rn\ and charged \galbh s, we find that the outcomes of the weak deflection lensings by the charged Horndeski and \galbh s are the same so that specific expressions for the charged \galbh\ are omitted. Meanwhile, the \rn\ and the tidal \rnbh s share the same formality but with opposite signs of the charges. Their observables of the weak deflection lensing are different from those of the charged Horndeski and \galbh s. However, these differences appear only in the next-to-leading-order terms. The details of weak deflection lensing by the (tidal) \rnbh\ are given in appendix \ref{app:rnwdl} (see also refs.\cite{Keeton2005PRD72.104006,Pang2019CQG36.065012}) and they are in the forms which can be directly compared with those of the charged Horndeski and \galbh s.

\subsection{Image positions and their relations}

According to the small parameter $\varepsilon$ in the weak deflection lensing, we assume that the position of a lensed image, which is a solution to the lens equation \eqref{eq:lenseq}, can be expressed in the form of a series as
\begin{equation}
  \label{}
  \theta = \theta_0 + \varepsilon \theta_1 + \varepsilon^2 \theta_2 + \mathcal{O}(\varepsilon^3),
\end{equation}
where $\theta_0$, $\theta_1$ and $\theta_2$ are respectively its 0th-, 1st- and 2nd-order approximations. Following the well-established procedure \cite{Keeton2005PRD72.104006,Keeton2006PRD73.044024,Keeton2006PRD73.104032}, we substitute it into the lens equation \eqref{eq:lenseq}, rearrange the lens equation in terms of $\varepsilon$ and obtain $\theta_n$  $(n=0,1,2)$ by vanishing the coefficients of $\varepsilon^n$. We find that $\theta_n$ for the charged \hornbh\ are
\begin{eqnarray}
  \label{}
  \theta_0 & = & \frac{1}{2}(\beta+\eta),\\
  \theta_1 & = & \frac{\pi}{4(\theta_0^2+1)}  \bigg(\frac{15}{4}-q\bigg),\\
  \theta_2 & = & \frac{1}{\theta_0 ( \theta_0^2 + 1 )^3} \bigg[ \frac{8}{3} D^2 \theta_0^8 + D \bigg( \frac{64}{3} D -16 \bigg) \theta_0^6 + \bigg( \frac{88}{3} D^2 - 32 D + 16 \bigg)  \theta_0^4 \nonumber\\
  & &  + \theta_0^2 \bigg( \frac{16}{3} D^2 - 16 D + 32 - \frac{225}{128} \pi^2 \bigg)  - \frac{16}{3} D^2 + 16 - \frac{225}{256} \pi^2  \bigg] \nonumber\\
  & & -\frac{q}{\theta_0 ( \theta_0^2 + 1 )^3} \bigg[ \frac{14}{3}\theta_0^4 + \bigg(\frac{28}{3}-\frac{15}{16}\pi^2 \bigg)\theta_0^2 -\frac{15}{32}\pi^2 + \frac{14}{3} \bigg] \nonumber\\
  & & -\frac{q^2\pi ( 2\theta_0^2 + 1 )}{16\theta_0 ( \theta_0^2 + 1 )^3},
\end{eqnarray}
where $D=D_{\mathrm{LS}}/D_{\mathrm{OS}}$ and
\begin{equation}
  \label{}
  \eta = \sqrt{\beta^2+4}.
\end{equation}
It is obvious that $\theta_0$ is not affected by the charge $q$ and its value is identical with the one of the \schbh\ in GR. The differences arise in $\theta_1$ and $\theta_2$ which return to the ones of the \schbh\ in the absence of $q$ \cite{Keeton2005PRD72.104006}.

In the aforementioned equations and hereafter, we adopt the convention of refs. \cite{Keeton2005PRD72.104006,Keeton2006PRD73.044024,Keeton2006PRD73.104032} that the angles of image positions are set to be positive. It means that the position of the source $\mathcal{B}$ is positive if the image is on the same side of the lens as the source, while it is negative if the image is on the opposite side. 

Therefore, we can respectively find out the positive- and negative-parity images at each orders as
\begin{eqnarray}
  \label{eq:theta0+-}
  \theta^{\pm}_0 & = & \frac{1}{2} ( \eta \pm |\beta| ), \\
  \label{eq:theta1+-}
  \theta^{\pm}_1 & = &  \frac{(15-4q)\pi }{8\eta (\eta \pm |\beta|)  }, \\
  \label{eq:theta2+-}
  \theta^{\pm}_2 & = & \frac{1}{\eta^3 (\eta\pm|\beta|)^4} \bigg\{ \frac{64}{3} D^2 \beta^8 + D \bigg( \frac{1024}{3} D - 128 \bigg) \beta^6  \nonumber\\
   & & + \bigg( \frac{5056}{3} D^2 - 1024 D + 128 \bigg) \beta^4  \nonumber\\
   & & + \bigg( \frac{8576}{3} D^2 - 2304 D + 768 - \frac{225}{16} \pi^2 \bigg) \beta^2  \nonumber\\
   & & + \frac{2560}{3} D^2 - 1024 D + 1024 - \frac{675}{16} \pi^2 \nonumber\\
  & &   -q \bigg[ \frac{112}{3} \beta^4 + \bigg(224 - \frac{15}{2} \pi^2\bigg) \beta^2+ \frac{896}{3} - \frac{45}{2} \pi^2 \bigg] \nonumber\\
  & & -q^2 \pi^2 (\beta^2+3) \bigg\} \nonumber\\  
  & & \pm \frac{\eta|\beta|}{\eta^3 (\eta\pm|\beta|)^4} \bigg[  \frac{64}{3} D^2 \beta^6 + D \bigg( \frac{896}{3} D - 128 \bigg)  \beta^4  \nonumber\\
  & &  +  \bigg( \frac{3392}{3} D^2 - 768 D + 128 \bigg)  \beta^2 \nonumber\\
  & &  + \frac{3328}{3} D^2 - 1024 D + 512 - \frac{225}{16} \pi^2 \nonumber\\
  & &   -q \bigg( \frac{112}{3} \beta^2 + \frac{448}{3} - \frac{15}{2} \pi^2 \big)  -q^2 \pi^2 \bigg],
\end{eqnarray}
which hold useful relations that
\begin{eqnarray}
  \label{}
  \theta^{+}_0 - \theta^{-}_0 & = & |\beta|, \\
  \label{}
  \theta^{+}_0  \theta^{-}_0 & = & 1, \\
  \label{}
  \theta^{+}_1 + \theta^{-}_1 & = & \frac{15-4q}{16}\pi, \\
  \label{}
  \theta^{+}_1 - \theta^{-}_1 & = & -\frac{(15-4q)\pi|\beta|}{16\eta},\\
  \label{eq:theta2+-theta2-}
  \theta^{+}_2 - \theta^{-}_2 & = & -|\beta| \bigg[ 16 - 8 D^2 - \frac{225}{256} \pi^2 \nonumber\\
  & & - q \bigg(\frac{14}{3} - \frac{15}{32} \pi^2 \bigg) - \frac{1}{16} \pi^2 q^2 \bigg].
\end{eqnarray}
It is clearly shown again that the differences between the charged Horndeski and the \schbh\ begin to appear in the next-to-leading-order approximations.

\subsection{Magnifications and their relations}

Using the same scheme, we expand $\mu$ into a series of $\varepsilon$ as
\begin{equation}
  \label{eq:mu}
  \mu = \mu_0 + \varepsilon \mu_1 + \varepsilon^2 \mu_2 + \mathcal{O}(\varepsilon^3),
\end{equation}
and obtain its 0th-, 1st- and 2nd-order approximations of the charged \hornbh\ as
\begin{eqnarray}
  \label{eq:mu0}
  \mu_0 & = & \frac{\theta_0^4}{\theta_0^4-1},\\
  \label{eq:mu1}
  \mu_1 & = & -\frac{(15-4q)\pi \theta_0^3}{16(\theta_0^2+1)^3},\\ 
  \label{eq:mu2}
  \mu_2 & = & \frac{\theta_0^2}{(\theta_0^2 + 1 )^5 (\theta_0^2 - 1)} \bigg[ \frac{8}{3} D^2 \theta_0^8 + (48 D^2 - 32 D - 32 ) \theta_0^6 \nonumber\\
  & & + \bigg( \frac{272}{3} D^2 - 64 D + \frac{675}{128} \pi^2 - 64 \bigg) \theta_0^4 \nonumber\\
  & &  + ( 48 D^2 - 32 D - 32 ) \theta_0^2 + \frac{8}{3} D^2\bigg] \nonumber\\
  & & + \frac{q\theta_0^4}{(\theta_0^2 + 1 )^5 (\theta_0^2 - 1)} \bigg[ \frac{28}{3}\theta_0^4 + \bigg( \frac{56}{3} - \frac{45}{16}\pi^2 \bigg) \theta_0^2 + \frac{28}{3}  \bigg] \nonumber\\
  & & + \frac{3q^2\pi^2\theta_0^6}{8(\theta_0^2 + 1 )^5 (\theta_0^2 - 1)}.
\end{eqnarray}
We further obtain their values for the positive- and negative-parity images as
\begin{eqnarray}
  \label{}
  \mu_0^{\pm} & = & \frac{1}{2}  \pm \frac{\beta^2+2}{2|\beta|\eta}, \\
  \label{}
  \mu_1^{+} & = & \mu_1^{-} =  -\frac{(15-4q) \pi}{16\eta^3} ,\\
  \label{eq:mu2+-}
  \mu^{\pm}_2 & = & \pm \frac{1}{|\beta|\eta^5} \bigg[\frac{8}{3} D^2 \beta^4 + \bigg(\frac{176}{3} D^2-32 D-32\bigg) \beta^2 \nonumber\\
  & & -128 D+192 D^2+ \frac{675}{128} \pi^2-128 \nonumber\\
  & & +q\bigg( \frac{28}{3}\beta^2+\frac{112}{3} - \frac{45}{16}\pi^2 \bigg)+\frac{3}{8}q^2\pi^2 \bigg],
\end{eqnarray}
which have very simple relations
\begin{eqnarray}
  \label{eq:mu0++mu0-}
  \mu_0^{+}+\mu_0^{-} & = & 1, \\
  \label{eq:mu1+-mu1-}
  \mu_1^{+}-\mu_1^{-} & = & 0, \\
  \label{eq:mu2++mu2-}
  \mu_2^{+}+\mu_2^{-} & = & 0.
\end{eqnarray}
and a clean combination 
\begin{equation}
  \label{}
  \mu_0^{+} \theta_1^{+} + \mu_0^{-} \theta_1^{-} + \mu_1^{+} \theta_0^{+} + \mu_1^{-}\theta_0^{-} = 0.
\end{equation}

Like the situations of the image positions and their relations, the leading term of the magnifications has no difference from its corresponding value for the \schbh\ while the elegant relations of the magnifications might be potentially used in the consistency check of observations on the lensed images.

\subsection{Total magnification and centroid}

If the two lensed images cannot be practically resolved, the observables will be the total magnification of them and their magnification-weighted centroid position. With the help of eqs. \eqref{eq:mu0++mu0-}--\eqref{eq:mu2++mu2-}, we can have the total magnification as
\begin{eqnarray}
  \label{eq:mutot}
  \mu_{\mathrm{tot}} & = & |\mu^+|+|\mu^-|\nonumber\\
  & = & (2\mu_0^+-1)+2\epsilon^2\mu_2^++\mathcal{O}(\varepsilon^3).
\end{eqnarray}
Since $\mu_1^+$ and $\mu_1^-$ cancel out each other exactly due to eq. \eqref{eq:mu1+-mu1-}, there is no $\mathcal{O}(\varepsilon)$ term, suggesting that distinguishing the charged \hornbh\ from the others by measuring total magnification requires the accuracy to reach the second-order level. 

The magnification-weighted centroid position is defined by 
\begin{equation}
  \label{}
  \Theta_{\mathrm{cent}} = \frac{\theta^{+}|\mu^{+}|-\theta^{-}|\mu^{-}|}{|\mu^{+}|+|\mu^{-}|} = \frac{\theta^{+}\mu^{+}+\theta^{-}\mu^{-}}{\mu^{+}-\mu^{-}},
\end{equation}
and it can be also expanded in the series of $\varepsilon$ as
\begin{equation}
  \label{}
  \Theta_{\mathrm{cent}} = \Theta_0+\varepsilon\Theta_1+\varepsilon^2\Theta_2+\mathcal{O}(\varepsilon^3),
\end{equation}
where the 0th-, 1st- and 2nd-order approximations are
\begin{eqnarray}
  \label{eq:Theta0}
  \Theta_0 & = & |\beta| \frac{\beta^2+3}{\beta^2+2},\\
  \label{eq:Theta1}
  \Theta_1 & = & 0,\\
  \label{eq:Theta2}
  \Theta_2 & = & \frac{|\beta|}{\eta^2(\beta^2 + 2) } \bigg[ \frac{8}{3} D^2 \beta^6  + \bigg( \frac{104}{3} D - 16 \bigg) D \beta^4 + \bigg( \frac{272}{3} D^2  \nonumber\\
  & & - 64 D + 32 \bigg) \beta^2 - \frac{64}{3} D^2- \frac{675}{128} \pi^2 + 128 \nonumber\\
  & & -q \bigg( \frac{28}{3}\beta^2 - \frac{45}{16}\pi^2 + \frac{112}{3} \bigg) - \frac{3}{8}q\pi^2 \bigg].
\end{eqnarray}
Likewise, the $\mathcal{O}(\varepsilon)$ term in $\Theta_{\mathrm{cent}}$ vanishes due to the cancellation between $\mu_1^+$ and $\mu_1^-$, demanding highly astrometric accuracy for any tests based on measuring the centroid position to access the second-order effects.

\subsection{Differential time delay}

The time delay is the difference between the light travel time with and without the lens and it can be expressed as \cite{Keeton2005PRD72.104006,Keeton2006PRD73.044024}
\begin{equation}
  \label{}
  c\tau = T(R_{\mathrm{src}}) + T(R_{\mathrm{obs}}) - \frac{D_{\mathrm{OS}}}{\cos\mathcal{B}}.
\end{equation}
The function $T(R)$, see eq. \eqref{eq:T(R)}, can be integrated and expanded as
\begin{equation}
  \label{eq:T(R)exp}
  T(R)  = T_0 + \sum_{k=1}^{3}\frac{m_{\bullet}^k}{r^k_0} r_0 T_k + \mathcal{O}\bigg(\frac{m_{\bullet}^4}{r_0^4}\bigg),
\end{equation}
where the functions $T_n$ in the 0th-, 1st-, 2nd- and 3rd-order approximations are
\begin{eqnarray}
  \label{}
  T_0 & = & \sqrt{R^2-r_0^2},\\
  T_1 & = & \frac{\sqrt{1-\xi^2}}{1+\xi} + 2 \ln\bigg(\frac{1+\sqrt{1-\xi^2}}{\xi}\bigg) ,\\
  T_2 & = & \bigg(\frac{15}{2}-2q\bigg)\bigg(\frac{\pi}{2}-\arcsin\xi\bigg) - \bigg(2+\frac{5}{2}\xi\bigg)\frac{\sqrt{1-\xi^2}}{(\xi+1)^2},\\
  T_3 & = & -\bigg(\frac{15}{2}-2q\bigg)\bigg(\frac{\pi}{2}-\arcsin\xi\bigg) \nonumber\\
  & & + \frac{\sqrt{1-\xi^2}}{2(\xi+1)^2} [ (35-18q)\xi^3 + (133-62q)\xi^2 \nonumber\\
  & & + (157-70 q) \xi + 60 -26q ],
\end{eqnarray}
with 
\begin{equation}
  \label{}
  \xi = \frac{r_0}{R}.
\end{equation}
The first term $T_0$ origins from the Euclidean geometry, while the second term is as the same as the Shapiro delay in GR which is not affected by the charged \hornbh.

Changing $r_0$ into $u$ and using the expressions of $R_{\mathrm{src}}$ and $R_{\mathrm{obs}}$, the scaled time delay can also be found as a series
\begin{equation}
  \label{}
  \hat{\tau} = \hat{\tau}_0 + \varepsilon \hat{\tau}_1 +\mathcal{O}(\varepsilon^2),
\end{equation}
where
\begin{eqnarray}
  \label{}
  \hat{\tau}_0 & = & \frac{1}{2} \bigg[1 + \beta^2 - \theta_0^2 - \ln \bigg(\frac{D_{\mathrm{OL}}\theta_0^2\vartheta_{\mathrm{E}}^2}{4D_{\mathrm{LS}}}\bigg)\bigg],\\
  \hat{\tau}_1 & = & \frac{(15-4q)\pi}{16\theta_0} .
\end{eqnarray}
In principle, the $\mathcal{O}(\varepsilon^2)$ term for $\hat{\tau}$ could be obtained as well, but it is less significant than the $\mathcal{O}(\varepsilon^2)$ corrections to the image positions $\theta$ and magnifications $\mu$. Once again, the leading effects of the time delay cannot tell difference of the charged \hornbh\ from the \schbh\ and detecting its next-to-leading-order term will be critical for that purpose. 

However, the absolute value of the time delay for a single lensed image is never accessible in the weak deflection lensing. The practical observable is the differential time delay between the positive- and negative-parity images as
\begin{equation}
  \label{}
  \Delta\hat{\tau}  =  \hat{\tau}_{-} - \hat{\tau}_{+}.
\end{equation}
It also has a series form
\begin{equation}
  \label{}
  \Delta\hat{\tau} = \Delta\hat{\tau}_0 + \varepsilon \Delta\hat{\tau}_1 + \mathcal{O}(\varepsilon^2),
\end{equation}
where
\begin{eqnarray}
  \label{}
  \Delta\hat{\tau}_0 & = &   \frac{1}{2} \eta |\beta| + \ln \bigg( \frac{\eta + |\beta|}{\eta - |\beta|} \bigg), \\
  \Delta\hat{\tau}_1 & = & \frac{15-4q}{16} \pi |\beta| 
\end{eqnarray}
It is theoretically feasible to test the charged \hornbh\ by measuring the differential time delay between two lensed images but such a measurement has to reach enough accuracy beyond the differential Shapiro delay.

\subsection{Practical observables}

Using these lensing quantities and their relations, we can have practical observables for the weak deflection lensing by the charged \hornbh. Therefore, the scaled variables $(\beta, \theta, \mu, \hat{\tau})$ must be converted to the physical ones $(\mathcal{B}, \vartheta, F,\tau)$ in which $F$ is the flux of the light. Observables of the weak deflection lensing usually are the positions, magnitudes of brightness, centroid and differential time delay between the lensed images. The fluxes are connected to the magnifications through the flux of the source, i.e., $F_i = |\mu_i| F_{\mathrm{src}}$. Some possibly measurable observables are \cite{Keeton2005PRD72.104006,Keeton2006PRD73.044024}
\begin{eqnarray}
  \label{eq:Ptot}
  P_{\mathrm{tot}} & \equiv & \vartheta^{+} + \vartheta^{-}  =  \mathcal{E} + \frac{15-4q}{16} \varepsilon \pi \vartheta_{\mathrm{E}}    +\mathcal{O}(\varepsilon^2), \\
  \label{}
  \Delta P & \equiv & \vartheta^{+} - \vartheta^{-}  = |\mathcal{B}| \bigg( 1 - \frac{15-4q}{16} \varepsilon \pi \frac{ \vartheta_{\mathrm{E}} }{ \mathcal{E} }  \bigg)  +\mathcal{O}(\varepsilon^2),\\
  \label{}
  F_{\mathrm{tot}} & \equiv & F^{+}+F^{-}  =  F_{\mathrm{src}} \frac{ \mathcal{B}^2 + 2 \vartheta_{\mathrm{E}}^2 } {|\mathcal{B}| \mathcal{E} } +\mathcal{O}(\varepsilon^2),\\
  \label{}
  \Delta F & \equiv & F^{+}-F^{-}  =  F_{\mathrm{src}}- F_{\mathrm{src}}\frac{15-4q}{8} \varepsilon \pi \frac{\vartheta_{\mathrm{E}}^3}{\mathcal{E}^{3}}  +\mathcal{O}(\varepsilon^2),\\
  \label{}
  S_{\mathrm{cent}} & \equiv & \frac{\vartheta^{+}F^{+}-\vartheta^{-}F^{-}}{F_{\mathrm{tot}}}  =  |\mathcal{B}| \frac{\mathcal{B}^2+3 \vartheta_{\mathrm{E}}^2}{\mathcal{B}^2+2 \vartheta_{\mathrm{E}}^2}  +\mathcal{O}(\varepsilon^2),\\
  \label{}
  \Delta \tau & = & \frac{D_{\mathrm{OL}}D_{\mathrm{OS}}}{cD_{\mathrm{LS}}} \bigg\{ \frac{1}{2} |\mathcal{B}| \mathcal{E}  +\vartheta_{\mathrm{E}}^2 \ln\Bigg( \frac{\mathcal{E}+|\mathcal{B}|}{\mathcal{E}-|\mathcal{B}|} \Bigg) + \varepsilon \frac{15-4q}{16} \pi \vartheta_{\mathrm{E}} |\mathcal{B}|   +\mathcal{O}(\varepsilon^2)\bigg\},
\end{eqnarray}
where
\begin{equation}
  \label{}
  \mathcal{E} = \sqrt{\mathcal{B}^2 + 4 \vartheta_{\mathrm{E}}^2 }.
\end{equation}
They are main signals of the weak deflection lensing by the charged \hornbh. Among them, the $\mathcal{O}(\varepsilon^2)$ terms in the practical observables are neglected because they are too small to be reachable in the foreseen future. Since the $\mathcal{O}(\varepsilon)$ terms of the total magnification \eqref{eq:mutot} and the centroid \eqref{eq:Theta1} are exactly cancelled, the total flux $F_{\mathrm{tot}}$ and the practical centroid $S_{\mathrm{cent}}$ of the lensed images have the $\mathcal{O}(1)$ terms only, which are not affected by any charge $q$. 

In order to indicate their deviations from those of the \schbh, we define that
\begin{eqnarray}
  \label{eq:defdPtot}
  \delta P_{\mathrm{tot}} & \equiv & P_{\mathrm{tot}} - P_{\mathrm{tot}}(q=0) = - \frac{q}{4} \varepsilon \pi \vartheta_{\mathrm{E}}    +\mathcal{O}(\varepsilon^2),\\
  \label{eq:defdDP}
  \delta \Delta P & \equiv & \Delta P - \Delta P(q=0) =  \frac{q}{4} \varepsilon \pi |\mathcal{B}| \frac{ \vartheta_{\mathrm{E}} }{ \mathcal{E} }  +\mathcal{O}(\varepsilon^2),\\
  \label{eq:drtot}
  \delta r_{\mathrm{tot}} & \equiv & 2.5 \log_{10} \bigg[ \frac{F_{\mathrm{tot}}}{F_{\mathrm{tot}}(q=0)} \bigg] = \mathcal{O}(\varepsilon^2),\\
  \label{eq:defdDr}
  \delta \Delta r & \equiv & 2.5 \log_{10} \bigg[\frac{\Delta F}{\Delta F(q=0)}\bigg] = \frac{5}{4\ln10}q\varepsilon \pi \frac{\vartheta_{\mathrm{E}}^3}{\mathcal{E}^{3}} +\mathcal{O}(\varepsilon^2),\\
  \label{eq:defdScent}
  \delta S_{\mathrm{cent}} & \equiv & F_{\mathrm{cent}} - F_{\mathrm{cent}}(q=0) =  \mathcal{O}(\varepsilon^2),\\
  \label{eq:defdDtau}
    \delta \Delta \tau & \equiv & \Delta \tau - \Delta \tau(q=0) =  -\frac{D_{\mathrm{OL}}D_{\mathrm{OS}}}{4cD_{\mathrm{LS}}}  \varepsilon q \pi \vartheta_{\mathrm{E}} |\mathcal{B}|   +\mathcal{O}(\varepsilon^2),
\end{eqnarray}
where the differences between fluxes are converted into magnitudes. The deviations in $\delta P_{\mathrm{tot}}$, $\delta\Delta P$, $\delta \Delta r$ and $\delta \Delta \tau$ are at the order of $\varepsilon$, making them likely to be detected in the near future, while the deviations in $\delta r_{\mathrm{tot}}$ and $\delta S_{\mathrm{cent}}$ are smaller than them by another of order $\varepsilon$, rendering them hardly to be detected.

Because Sgr A* is the only supermassive black hole with orbiting stars that can be directly observed, we consider it as the lens with $M=4.28\times 10^6$ $M_{\odot}$ and $D_{\mathrm{OL}}=8.32$ kpc \cite{Gillessen2017ApJ837.30} and a source orbiting it with a distance $D_{\mathrm{LS}}=10^{-3}$ pc whose the angular Einstein radius is $\theta_{\mathrm{E}}=710$ $\mu$as and small parameter $\varepsilon$ is $7.2\times10^{-3}$ due to $D_{\mathrm{OS}}\approx D_{\mathrm{OL}}$. We note that the star S175 orbiting Sgr A* has the periastron distance of $2\times10^{-4}$ pc \cite{Gillessen2017ApJ837.30}, less than our assumption of $D_{\mathrm{LS}}$. We will discuss $P_{\mathrm{tot}}$, $\Delta P$, $\Delta F$ and $\Delta \tau$ and their deviations from those of the \schbh, since $q$ appears at their first order approximations in terms of $\varepsilon$. 

\subsection{Example for Sgr A*}

\begin{figure}[phtb]
\centering 
\includegraphics[width=.75\textwidth]{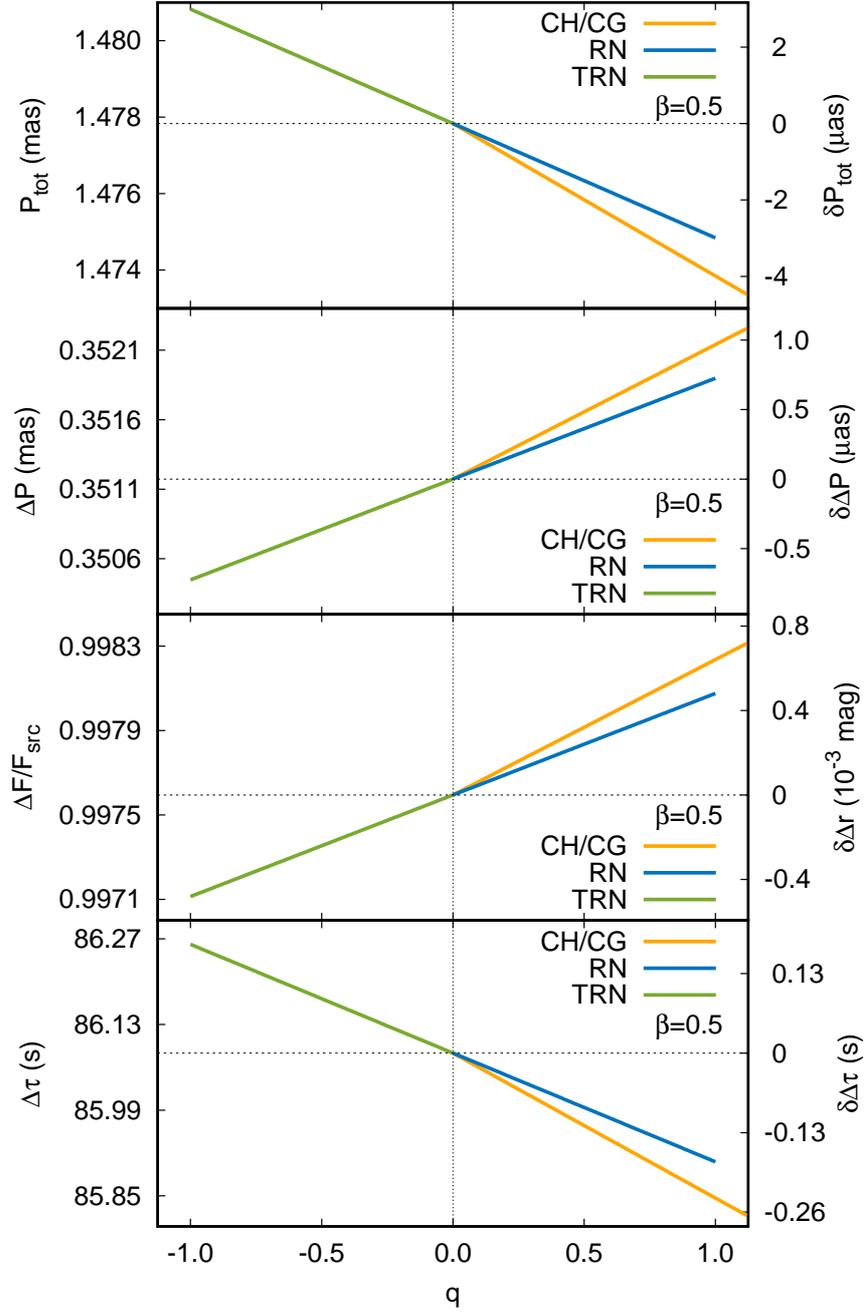}
\caption{\label{fig:WDLSgrA} Weak deflection lensing observables (left $y$-axis) and their deviations from those of the \schbh\ in GR (right $y$-axis) by Sgr A* in the cases of the charged Horndeski (CH), \rn\ (RN), tidal \rn\ (TRN) and charged Galileon (CG) black holes for $\beta=0.5$.}
\end{figure}

Figure \ref{fig:WDLSgrA} shows these lensing observables (left $y$-axis) and their deviations from those of the \schbh\ (right $y$-axis) by the charged Horndeski, \rn, tidal \rn\ and charged \galbh s with Sgr A* as the lens for $\beta=0.5$. It is worth mentioning that the charged Horndeski and the charged \galbh s have the same observables for the weak deflection lensing and the \rn\ and tidal \rnbh s have the same formulae for their weak deflection lensings and observables but with opposite signed $q$ (see appendix \ref{app:rnwdl}) .

For all of these black holes, the angular separations between two lensed images $P_{\mathrm{tot}}$ are about $1.47$ milliarcsecond (mas), much larger than the current resolution of 50 $\mu$as realized by phase referencing optical/infrared interferometry \cite{GRAVITYCol2017A&A602.A94} so that these two images would potentially be able to resolved. For a given positive $q$, $P_{\mathrm{tot}}$ of the \rnbh\ is slightly bigger than the one of the charged \hornbh, while the negative $q$ makes the tidal \rnbh\ has even larger $P_{\mathrm{tot}}$. Their deviations from the one of the \schbh\ $\delta P_{\mathrm{tot}}$ range from $-4$ to $2$ $\mu$as, where the deviation of the tidal \rnbh\ is positive and the others are negative. Although the $\delta P_{\mathrm{tot}}$ of the charged \hornbh\ is most significant among them, it is too small to detect for current technology. The angular differences between the two lensed images $\Delta P$ for these black holes are all about $0.3$ mas, potentially resolvable for current ability. Contrary to the cases of $P_{\mathrm{tot}}$, the charged \hornbh\ has the biggest $\Delta P$ and the tidal \rnbh\ has the smallest one. Their deviations from the one of the \schbh\ $\delta\Delta P$ are mostly below the level of 1 $\mu$as, making them undetectable for now. The normalized fluxes difference $\Delta F/F_{\mathrm{src}}$ for these black holes are close to $0.998$ and their absolute deviations from the \schbh's $\delta \Delta r$ are about $4\times10^{-4}$ mag, equivalent to relative flux of about 370 parts per million (ppm), where the charged \hornbh\ has the biggest values of these two quantities for a given $|q|$. Such a small difference of $\delta \Delta r$ is (marginally) reachable for a space telescope, such as Transiting Exoplanet Survey Satellite with 300 ppm photometric resolution \cite{Huang2018ApJ868.L39}, but flares of Sgr A* would wipe it out easily. The differential time delay between the two lensed images $\Delta \tau$ for these black holes are at the level of $86$ s and their absolute deviations from the \schbh's $\delta\Delta \tau$ are less than 0.26 s. Among them, the charged \hornbh\ has the smallest values of $\Delta \tau$ and $\delta\Delta \tau$ for a given $|q|$. These two timing signals are both significantly shorter than the time span of a typical astronomical observation session lasting for hours and, therefore, are unable to be measured. 

In a summary of the weak deflection lensing by these charged black holes, we find that (1) among the lensing observables, the leading effects of $P_{\mathrm{tot}}$, $\Delta P$ and $\Delta F$ can be detected by current technology; (2) among their deviations from those of the \schbh, only $\delta \Delta r$ is within today's capability if the flare of Sgr A* can be well handled; and (3) the existing technology is unable to detect the charge of these black holes and distinguish them further based on the deviations of their lensing observables from those of the \schbh.

\section{Strong deflection lensing}

\label{sec:sdl}

In the scenario of strong deflection lensing, the closet approach distance of a light ray to the lens is very close to its gravitational radius so that the bending angle \eqref{eq:dfagl} will increase and eventually diverge as $r_0$ decreases. It suggests that, before photons in the ray arrive the observer, they wind around the lens for several loops, which never happens in the weak deflection lensing.

\subsection{Strong deflection limit and observables}

An effectively analytic way to deal with such a divergence is to expand the integral of the bending angle near the photon sphere by the method of strong deflection limit \cite{Bozza2002PRD66.103001}. The photon sphere is the innermost circular orbit for a photon and its radius satisfies the following equation \cite{Virbhadra2000PRD62.084003,Claudel2001JMP42.818}
\begin{equation}
\label{eq:phosph}
\frac{C'(r)}{C(r)}=\frac{A'(r)}{A(r)},
\end{equation}
where $'$ denotes derivative against $r$. It is the biggest root of the resulting quartic equation of $r$, which can be found as $r_m$. Therefore, the deflection angle in the strong deflection lensing can be expanded as \cite{Bozza2002PRD66.103001}
\begin{equation}
  \label{eq:SDLalpha}
  \hat{\alpha}(\theta) = -\bar{a}\log\left(\frac{\theta D_{OL}}{u_m}-1\right)+\bar{b}+\mathcal{O}[(u-u_m)\log(u-u_m)],
\end{equation}
where $u$ is the impact parameter satisfying $C(r_0)=u^2 A(r_0)$ and $r_0$ is the closet approach distance of the photon. The subscript $m$ means evaluating at $r=r_m$ at the photon sphere so that $u_m$ holds $C(r_m)=u_m^2 A(r_m)$. The strong deflection limit coefficients in the above equation can be obtained as \cite{Bozza2002PRD66.103001}
\begin{eqnarray}
  \bar{a} & = & \frac{R_m}{2\sqrt{\gamma_m}},\\
  \bar{b} & = & -\pi + b_R+\bar{a}\ln\bigg(\frac{2\gamma_m}{A_m}\bigg),
\end{eqnarray}
where some immediate quantities are
\begin{eqnarray}
  \gamma_m & = & {C_m(1-A_m)^2 \left(A_m C_m''-C_m A_m''\right)\over{2A_m^2C_m'^2}},\\
   R_m & = & \frac{2(1-A_m)\sqrt{A_mB_m}}{A'_m\sqrt{C_m}},\\
   b_R & = & \int^1_0 \left[\frac{2(1-A_m)\sqrt{A(z)B(z)}}{A'(z)C(z)\sqrt{\frac{A_m}{C_m}-\frac{A(z)}{C(z)}}}-\frac{R_m}{z\sqrt{\gamma_m}}\right]\mathrm{d}z,
\end{eqnarray}
and the variable $z$ in the integral is defined by
\begin{equation}
\label{eq:xtoz}
  z = \frac{A(r)-A_m}{1-A_m},
\end{equation}
with $'$ and $''$ meaning derivative against $r$ once and twice. They ensure that $\bar{a}$ and $\bar{b}$ can be calculated according to the metric (\ref{eq:metric}) of the charged \hornbh\ for any valid $q$. If we consider that both the source and the observer are very far from the lens and in the asymptotically flat spacetime and three of them are nearly aligned with the lens, then the lens equation (\ref{eq:lenseq}) can be simplified as \cite{Bozza2001GRG33.1535}
\begin{equation}
\label{eq:SDLLE}
\beta=\theta-\frac{D_{\mathrm{LS}}}{D_{\mathrm{OS}}}[\hat{\alpha}(\theta)-2n\pi],\qquad n\in \mathbb{Z}
\end{equation}
where $\beta$, $\theta$ and $\hat{\alpha}(\theta)-2n\pi$ are all very small. The integer $n$ denotes the number of loops wound around by the photon, which will form the resulting relativistic image.

In order to study the differential time delay between two relativistic images of a variable source, we change the total time span, eq. (\ref{eq:T}), into the following form as \cite{Bozza2004GRG36.435}
\begin{equation}
  \label{}
  T=\tilde{T}(r_0)-\int^\infty_{D_{\mathrm{OL}}}\left|\frac{\mathrm{d}t}{\mathrm{d}r}\right|\mathrm{d}r-\int^\infty_{D_{\mathrm{LS}}}\left|\frac{\mathrm{d}t}{\mathrm{d}r}\right|\mathrm{d}r.
\end{equation}
Based on the facts that the source and the observer are in the asymptotically flat spacetime, we can obtain its second and third terms by means of the approach in the weak deflection lensing, while the first term coming from the strong deflection lensing is 
\begin{equation}
\label{eq:dftime}
\tilde{T}(r_0)=\int^{\infty}_{r_0}2\left|\frac{\mathrm{d}t}{\mathrm{d}r}\right|\mathrm{d}r,
\end{equation}
where $\mathrm{d}t/\mathrm{d}r$ can be found in eq. \eqref{eq:dtdr}. It can also be integrated in the strong deflection limit as \cite{Bozza2004GRG36.435}
\begin{equation}
\label{eq:SDLtime}
\tilde{T}(u)=-\tilde{a}\ln{\left(\frac{u}{u_m}-1\right)}+\tilde{b}+\mathcal{O}[(u-u_m)\log(u-u_m)],
\end{equation}
where $\tilde{a}$ and $\tilde{b}$ are strong deflection limit coefficients and $\tilde{a}= \bar{a}\, u_m$ for the charged \hornbh\ (\ref{eq:metric}).

With the helps of the deflection angle (\ref{eq:SDLalpha}) and the lens equation (\ref{eq:SDLLE}), the apparent radius of the photon sphere $\theta_\infty$, the angular separation between the first relativistic image and other packed images $s$ and their brightness difference $ \Delta m$ can be found as \cite{Bozza2002PRD66.103001}
\begin{eqnarray}
  \label{eq:theta}
  \theta _{\infty } & = & \frac{u_m}{D_{\mathrm{OL}}},\\
  \label{eq:s}
  s & = & \theta _{\infty } \exp \left(\frac{\bar{b}}{\bar{a}}-\frac{2 \pi }{\bar{a}}\right),\\
  \label{eq:r}
  \Delta m & = & 2.5 \log_{10}\bigg[\exp \bigg(\frac{2 \pi }{\bar{a}}\bigg)\bigg].
\end{eqnarray}
If the brightness of the source varies,  we can also calculate the differential delay between the first and second relativistic images $\Delta T_{2,1}$ by making use of eq. (\ref{eq:SDLtime}) as \cite{Bozza2004GRG36.435}
\begin{equation}
\label{eq:TD}
\Delta T_{2,1} = \Delta T_{2,1}^{0}+\Delta T_{2,1}^{1},
\end{equation}
where the leading and correction terms are
\begin{eqnarray}
  \label{eq:TD0}
  \Delta T_{2,1}^{0} & = & 2 \pi  u_m ,\\
  \label{eq:TD1}
  \Delta T_{2,1}^{1} & = & 2\sqrt{\frac{B_m}{A_m}}\sqrt{\frac{u_m}{c_m}}\exp\left({\frac{\bar{b}}{\bar{a}}}\right) \left[\exp\left({-\frac{\pi}{\bar{a}}}\right)-\exp\left({-\frac{2\pi}{\bar{a}}}\right)\right]
\end{eqnarray}
with
\begin{equation}
\label{eq:cm}
c_m=\gamma_m\sqrt{\frac{A_m}{C_m^3}}\frac{{C_m'}{}^2}{2(1-A_m)^2}
\end{equation}
and their ratio
\begin{equation}
  \label{}
  \eta_{2,1} = \frac{\Delta T_{2,1}^{1}}{\Delta T_{2,1}}.
\end{equation}
Since $\Delta T_{2,1}^{0}$ is effectively equivalent to $u_{m}$, only $\Delta T_{2,1}^{1}$ can provide extra information.

The effects of the charged \hornbh\ with respect to their corresponding values of the \schbh\ in GR can also be indicated by
\begin{eqnarray}
  \label{}
  \delta \theta_{\infty} & = &  \theta_{\infty} - \theta_{\infty}(q=0),\\
  \delta s & = & s - s(q=0),\\
  \delta  \Delta m & = &  \Delta m -  \Delta m(q=0),\\
  \delta \Delta T_{2,1} & = &  \Delta T_{2,1} - \Delta T_{2,1}(q=0) ,\\
  \delta \eta_{2,1} & = & \eta_{2,1} - \eta_{2,1}(q=0).
\end{eqnarray}

\subsection{Example for Sgr A*}

\begin{figure}[tp]
\centering 
\includegraphics[width=.65\textwidth]{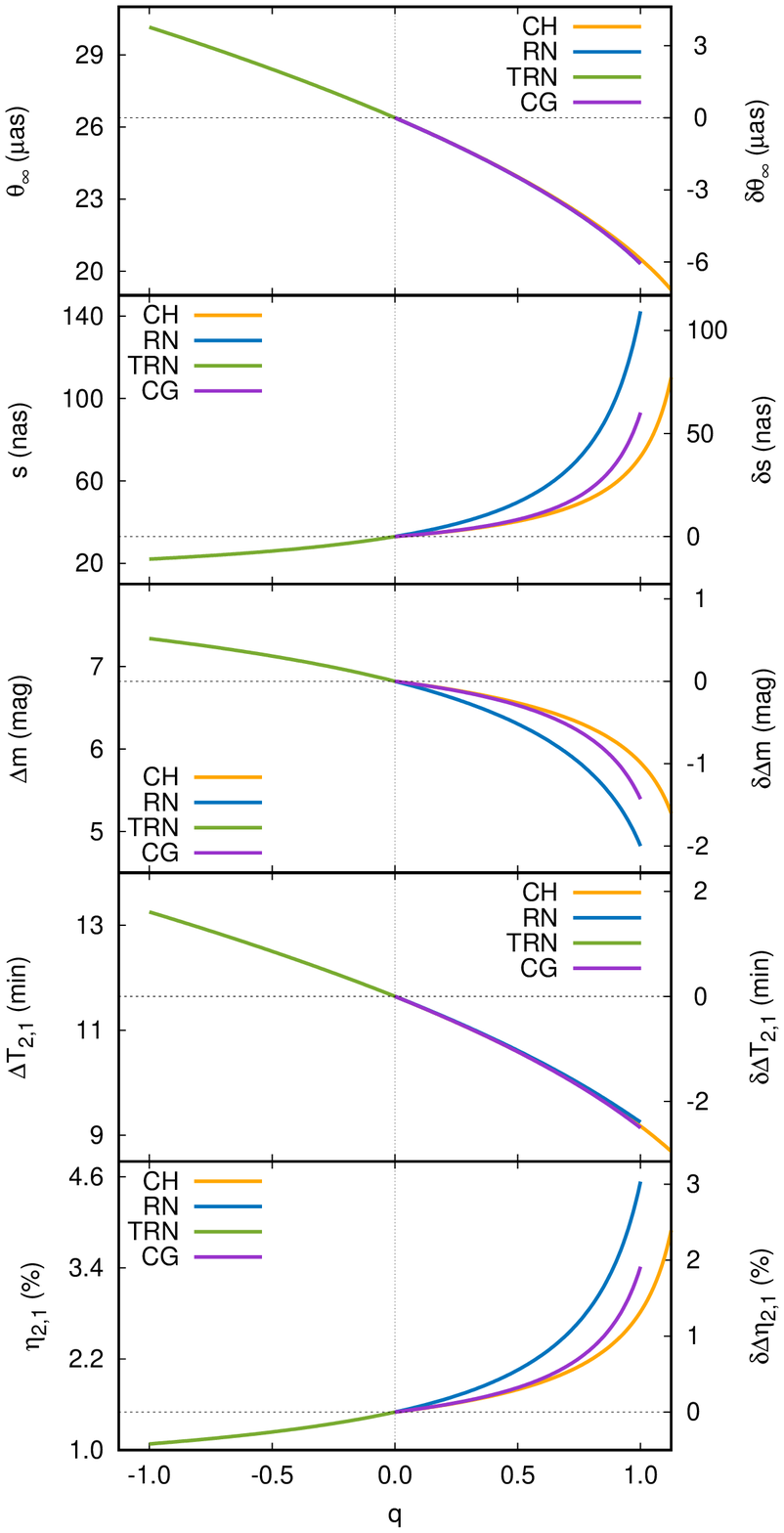}
\caption{\label{fig:SDLSgrA} Strong deflection lensing observables (left $y$-axis) and their deviations from those of the \schbh\ in GR (right $y$-axis) by Sgr A* in the cases of the charged \hornbh\ (CH), \rn\ (RN), tidal \rn\ (TRN) and charged Galileon (CG) black holes.}
\end{figure}

Figure \ref{fig:SDLSgrA} shows strong deflection lensing observables (left $y$-axis) and their deviations from those of the \schbh\ (right $y$-axis) by the charged Horndeski, \rn, tidal \rn\ and charged \galbh s with Sgr A* as the lens. The strong deflection lensing of the \rn, tidal \rn\ and charged \galbh\ were respectively investigated in refs. \cite{Bozza2002PRD66.103001,Eiroa2002PRD66.024010,Zakharov2005AA442.795,Pang2019CQG36.065012}, refs. \cite{Bin-Nun2010PRD81.123011,Bin-Nun2010PRD82.064009,Horvath2013AN334.1047,Whisker2005PRD71.064004,Zakharov2014PRD90.062007} and ref. \cite{Zhao2016JCAP07.007}. 

For all of these black holes, their apparent radius of the photon sphere (``shadow'') $\theta_{\infty}$ range from $19$ to $30$ $\mu$as, which would be resolved by EHT with angular resolution of 20 $\mu$as \cite{Doeleman2008Nature455.78}. The negative charge $q$ can enlarge $\theta_{\infty}$ for the tidal \rnbh, while the positive charge $q$ make it shrink for the other black holes, which have almost the same curves of $\theta_{\infty}$ against $q$. However, since the charged \hornbh\ can have a larger $q$ up to $9/8$, it can possess an even smaller shadow. Whereas EHT can resolve the shadow, it cannot distinguish these black holes because the absolute deviations of their $\theta_{\infty}$ from the one of the \schbh\ are no more than $7$ $\mu$as. The angular separations between the first relativistic image and other packed images $s$ for these black holes change significantly from 20 to 150 nanosecond (nas). Compared with the one of the \schbh, the negative charge of the tidal \rnbh\ suppress this separation, while the other black holes with positive charges can magnify it by a factor of $4$ and have obviously different dependence on $q$. For a given positive $q$, the charged \hornbh\ has the smallest values of $s$ and its behaviour on $s$ is more similar to the charged \galbh's than the \rnbh's. Although this situation benefits telling one kind of black hole from the others, the tininess of $s$ makes it hardly accessible in the foreseen future. Contrary to the cases of $s$ for these black holes, the brightness differences between the first relativistic image and other packed images $\Delta m$ show opposite relations on $q$. The negative charge of the tidal \rnbh\ can raise the contrast of them, while the positive charges of other black holes make the brightness of them closer. The magnitude of $\Delta m$ is theoretically within the current ability of photometry but measuring it is practically impossible because there is not enough resolution to separate these relativistic images. The differential time delay between the relativistic images $\Delta T_{2,1}$ for these black holes are all less than $14$ min, where the tidal \rnbh\ has the biggest values and others are smaller than the one of the \schbh\ by about $3$ min. The curves of $\Delta T_{2,1}$ have almost the same shapes with those of $\theta_{\infty}$, which is caused by the dominance of $\Delta T_{2,1}^0\propto u_m$ in $\Delta T_{2,1}$ as the ratio of its correction to the total delay $\eta_{2,1}$ is less than $5\%$. $\Delta T_{2,1}$ of Sgr A* is unable to measure since it is much shorter than the time span for observation sessions of EHT.

In a summary of the strong deflection lensing of these charged black holes, we find that (1) among the lensing observables, the apparent size of the photon sphere for Sgr A* is within the current capability of EHT; (2) there is not enough resolution to distinguish these kinds of black holes based on such a measurement; and (3) resolving the relativistic images of these black holes requires technology far beyond this age so that it would be impossible to measure their angular separation, brightness difference and differential time delay in the foreseen future.

\section{Conclusions and discussion}

\label{sec:con}

We investigate the weak and strong deflection gravitational lensings by the charged \hornbh\ and compare its signals with those of the \rn, tidal \rn\ and charged \galbh s, hoping to provide hints for distinguishing these kind of black holes by observing the supermassive black hole in the Galactic Center, Sgr A*, with infrared and radio interferometry in the future. Weak deflection lensing observables for the charged Horndeski and other black holes, including positions, magnifications and differential time delay between the lensed images and their relations are obtained. The practical observables are also constructed and analysed. We find that, for Sgr A* as the lens, although the angular separation, angular difference and fluxes difference between the two lensed images are within the thresholds of current technology, the deviations of these observables from those of the \schbh\ are either too small to detect or easily to be wiped out by the flares of Sgr A*. Strong deflection lensing observables, including the apparent radius of the photon sphere as well as the angular separation, brightness difference and differential time delay between the relativistic images, are found and evaluated by taking Sgr A* as the lens. We find that only the apparent size of the photon sphere can be potentially measured by EHT whereas its resolution is not enough to distinguish these charged black holes, and there is no technology that can resolve the relativistic images of them in the foreseen future. 

The charged \hornbh\ we have discussed is a non-rotating one. Nevertheless, an astrophysical black hole is very likely spinning. This fact might make our result (partially) inapplicable if this spin is not negligible. In order to describe a rotating charged \hornbh\ and obtain its weak and strong lensing signals, we must have its metric, which is, to our knowledge, unavailable for now. Based on previous works on light propagation and gravitational lensing by a charged and/or spinning black hole in its weak \cite{Ibanez1983A&A124.175,Bray1986PRD34.367,Klioner1991SvA35.523,Glicenstein1999A&A343.1025,Sereno2006PRD74.123009,Werner2007PRD76.064024,Sereno2008PRD78.023008,Aazami2011JMP52.092502,Aazami2011JMP52.102501,He2014IJMPD23.1450031,He2014IJMPD23.1450079,Deng2015IJMPD24.1550056,He2015RAA15.646,Deng2016IJMPD25.1650082,He2016PRD93.023005,He2016PRD94.063011,He2017CQG34.105006} and strong \cite{Bardeen1973BH.215,Bozza2003PRD67.103006,Vazquez2004NCimB119.489,Bozza2005PRD72.083003,Bozza2006PRD74.063001,Bozza2008PRD78.063014,Iyer2009PRD80.124023} fields, we would expect that the spinning of an charged \hornbh\ would shift the caustic and extend it to a finite shape, distort and displace the photon sphere, and make the lens equation more complex, which will be left for future studies. The determinations of the charge and the spacetime of Sgr A* based on observations are also complicated works. Emissions and flares of Sgr A* can make the lensed images in the weak deflection lensing hardly resolvable, while plasma in the environments of Sgr A*  and its underlying general relativistic magnetohydrodynamics can render the apparent shape of the photon sphere no longer regular. Therefore, we just give direct hints of weak and strong deflection gravitational lensing signals by a non-rotating charged \hornbh\ with analytical methods. It indeed requires more sophisticated researches in the future.

\appendix

\section{Weak deflection lensing by a (tidal) \rnbh}

\label{app:rnwdl}

Since the \rn\ and tidal \rnbh s have the same metric but with opposite charge $q$, we only show the results of the \rn\ one here. In the weak deflection lensing, the deflection angle for a light ray caused by the \rnbh\ is
\begin{equation}
  \label{eq:RNhatalphab}
  \hat{\alpha}_{\mathrm{RN}}(u) = 4\frac{m_{\bullet}}{u} + \frac{3}{4} (5-q) \pi \bigg(\frac{m_{\bullet}}{u}\bigg)^2  + 16 \bigg(\frac{8}{3} -  q\bigg) \bigg(\frac{m_{\bullet}}{u}\bigg)^3 +\mathcal{O}\bigg(\frac{m_{\bullet}^4}{u^4}\bigg).
\end{equation}
The position of its lensed image is
\begin{equation}
  \label{}
  \theta_{\mathrm{RN}} = \theta_{\mathrm{RN}0} + \varepsilon \theta_{\mathrm{RN}1} + \varepsilon^2 \theta_{\mathrm{RN}2} + \mathcal{O}(\varepsilon^3),
\end{equation}
where
\begin{eqnarray}
  \label{}
  \theta_{\mathrm{RN}0} & = & \frac{1}{2}(\beta+\eta),\\
  \theta_{\mathrm{RN}1} & = & \frac{3(5-q)\pi}{16(\theta_0^2+1)} ,\\
  \theta_{\mathrm{RN}2} & = & \frac{1}{\theta_0 ( \theta_0^2 + 1 )^3} \bigg[ \frac{8}{3} D^2 \theta_0^8 + D \bigg( \frac{64}{3} D -16 \bigg) \theta_0^6 + \bigg( \frac{88}{3} D^2 - 32 D + 16 \bigg)  \theta_0^4 \nonumber\\
  & &  + \theta_0^2 \bigg( \frac{16}{3} D^2 - 16 D + 32 - \frac{225}{128} \pi^2 \bigg)  - \frac{16}{3} D^2 + 16 - \frac{225}{256} \pi^2  \bigg] \nonumber\\
  & & -\frac{q}{\theta_0 ( \theta_0^2 + 1 )^3} \bigg[ 4\theta_0^4 + \bigg( 8-\frac{45}{64}\pi^2 \bigg)\theta_0^2 -\frac{45}{128}\pi + 4 \bigg] \nonumber\\
  & & -\frac{9 q^2\pi ( 2\theta_0^2 + 1 )}{256 \theta_0 ( \theta_0^2 + 1 )^3}.
\end{eqnarray}
Therefore, the positions of the positive- and negative-parity images for the \rnbh\ are
\begin{eqnarray}
  \label{}
  \theta^{\pm}_{\mathrm{RN}0} & = & \frac{1}{2} ( \eta \pm |\beta| ), \\
  \theta^{\pm}_{\mathrm{RN}1} & = & \frac{3(5-q)\pi }{8\eta (\eta \pm |\beta|)  } ,\\
  \label{eq:RNtheta2+-}
  \theta^{\pm}_{\mathrm{RN}2} & = & \frac{1}{\eta^3 (\eta\pm|\beta|)^4} \bigg\{ \frac{64}{3} D^2 \beta^8 + D \bigg( \frac{1024}{3} D - 128 \bigg) \beta^6  \nonumber\\
   & & + \bigg( \frac{5056}{3} D^2 - 1024 D + 128 \bigg) \beta^4  \nonumber\\
   & & + \bigg( \frac{8576}{3} D^2 - 2304 D + 768 - \frac{225}{16} \pi^2 \bigg) \beta^2  \nonumber\\
   & & + \frac{2560}{3} D^2 - 1024 D + 1024 - \frac{675}{16} \pi^2 \nonumber\\
  & &   -q \bigg[ 32 \beta^4 + \bigg(192 - \frac{45}{8} \pi^2\bigg) \beta^2+ 256 - \frac{135}{8} \pi^2 \bigg] \nonumber\\
  & & - \frac{9}{16} q^2 \pi^2 (\beta^2+3) \bigg\} \nonumber\\  
  & & \pm \frac{\eta|\beta|}{\eta^3 (\eta\pm|\beta|)^4} \bigg[  \frac{64}{3} D^2 \beta^6 + D \bigg( \frac{896}{3} D - 128 \bigg)  \beta^4  \nonumber\\
  & &  +  \bigg( \frac{3392}{3} D^2 - 768 D + 128 \bigg)  \beta^2 \nonumber\\
  & &  + \frac{3328}{3} D^2 - 1024 D + 512 - \frac{225}{16} \pi^2 \nonumber\\
  & &   -q \bigg( 32 \beta^2 + 128 - \frac{45}{8} \pi^2 \big)  - \frac{q^2}{16} \pi^2 \bigg].
\end{eqnarray}
They have relations as  
\begin{eqnarray}
  \label{}
  \theta^{+}_{\mathrm{RN}0} - \theta^{-}_{\mathrm{RN}0} & = & |\beta|, \\
  \label{}
  \theta^{+}_{\mathrm{RN}0}  \theta^{-}_{\mathrm{RN}0} & = & 1, \\
  \label{}
  \theta^{+}_{\mathrm{RN}1} + \theta^{-}_{\mathrm{RN}1} & = & \frac{3(5-q)}{16}\pi, \\
  \label{}
  \theta^{+}_{\mathrm{RN}1} - \theta^{-}_{\mathrm{RN}1} & = & -\frac{3(5-q)\pi|\beta|}{16\eta},\\
  \label{eq:theta2+-theta2-}
  \theta^{+}_{\mathrm{RN}2} - \theta^{-}_{\mathrm{RN}2} & = & -|\beta| \bigg[ 16 - 8 D^2 - \frac{225}{256} \pi^2 \nonumber\\
  & & - q \bigg(4 - \frac{45}{128} \pi^2 \bigg) - \frac{9}{256} q^2 \pi^2  \bigg],
\end{eqnarray}

The magnifications of the weak deflection lensing by the \rnbh\ are
\begin{equation}
  \label{eq:mu}
  \mu_{\mathrm{RN}} = \mu_{\mathrm{RN}0} + \varepsilon \mu_{\mathrm{RN}1} + \varepsilon^2 \mu_{\mathrm{RN}2} + \mathcal{O}(\varepsilon^3),
\end{equation}
where
\begin{eqnarray}
  \label{eq:RNmu0}
  \mu_{\mathrm{RN}0} & = & \frac{\theta_0^4}{\theta_0^4-1},\\
  \label{eq:RNmu1}
  \mu_{\mathrm{RN}1} & = & -\frac{3(5-q)\pi \theta_0^3}{16(\theta_0^2+1)^3},\\ 
  \label{eq:RNmu2}
  \mu_{\mathrm{RN}2} & = & \frac{\theta_0^2}{(\theta_0^2 + 1 )^5 (\theta_0^2 - 1)} \bigg[ \frac{8}{3} D^2 \theta_0^8 + (48 D^2 - 32 D - 32 ) \theta_0^6 \nonumber\\
  & & + \bigg( \frac{272}{3} D^2 - 64 D + \frac{675}{128} \pi^2 - 64 \bigg) \theta_0^4 \nonumber\\
  & &  + ( 48 D^2 - 32 D - 32 ) \theta_0^2 + \frac{8}{3} D^2\bigg] \nonumber\\
  & & + \frac{q\theta_0^4}{(\theta_0^2 + 1 )^5 (\theta_0^2 - 1)} \bigg[ 8\theta_0^4 + \bigg( 16 - \frac{135}{64}\pi^2 \bigg) \theta_0^2 + 8  \bigg] \nonumber\\
  & & + \frac{27q^2\pi^2\theta_0^6}{128(\theta_0^2 + 1 )^5 (\theta_0^2 - 1)}.
\end{eqnarray}
Their values for the positive- and negative-parity images are
\begin{eqnarray}
  \label{}
  \mu_{\mathrm{RN}0}^{\pm} & = & \frac{1}{2}  \pm \frac{\beta^2+2}{2|\beta|\eta}, \\
  \label{}
  \mu_{\mathrm{RN}1}^{+} & = & \mu_{\mathrm{RN}1}^{-} =  -\frac{3(5-q) \pi}{16\eta^3} ,\\
  \label{eq:RNmu2+-}
  \mu^{\pm}_{\mathrm{RN}2} & = & \pm \frac{1}{|\beta|\eta^5} \bigg[\frac{8}{3} D^2 \beta^4 + \bigg(\frac{176}{3} D^2-32 D-32\bigg) \beta^2 \nonumber\\
  & & -128 D+192 D^2+ \frac{675}{128} \pi^2-128 \nonumber\\
  & & +q\bigg( 8\beta^2 + 32 - \frac{135}{64}\pi^2 \bigg) + \frac{27}{128}q^2\pi^2 \bigg],
\end{eqnarray}
which also hold relations as
\begin{eqnarray}
  \label{eq:RNmu0++RNmu0-}
  \mu_{\mathrm{RN}0}^{+}+\mu_{\mathrm{RN}0}^{-} & = & 1, \\
  \label{eq:RNmu1+-RNmu1-}
  \mu_{\mathrm{RN}1}^{+}-\mu_{\mathrm{RN}1}^{-} & = & 0, \\
  \label{eq:RNmu2++RNmu2-}
  \mu_{\mathrm{RN}2}^{+}+\mu_{\mathrm{RN}2}^{-} & = & 0,
\end{eqnarray}
and  
\begin{equation}
  \label{}
  \mu_{\mathrm{RN}0}^{+} \theta_{\mathrm{RN}1}^{+} + \mu_{\mathrm{RN}0}^{-} \theta_{\mathrm{RN}1}^{-} + \mu_{\mathrm{RN}1}^{+} \theta_{\mathrm{RN}0}^{+} + \mu_{\mathrm{RN}1}^{-}\theta_0^{-} = 0.
\end{equation}

The total magnification can be found as
\begin{eqnarray}
  \label{}
  \mu_{\mathrm{RN,tot}}  =   (2\mu_{\mathrm{RN}0}^+-1)+2\epsilon^2\mu_{\mathrm{RN}2}^++\mathcal{O}(\varepsilon^3),
\end{eqnarray}
while the position of the centroid is  
\begin{equation}
  \label{}
  \Theta_{\mathrm{RN,cent}} = \Theta_{\mathrm{RN}0}+\varepsilon\Theta_{\mathrm{RN}1}+\varepsilon^2\Theta_{\mathrm{RN}2}+\mathcal{O}(\varepsilon^3),
\end{equation}
where
\begin{eqnarray}
  \label{}
  \Theta_{\mathrm{RN}0} & = & |\beta| \frac{\beta^2+3}{\beta^2+2},\\
  \label{}
  \Theta_{\mathrm{RN}1} & = & 0,\\
  \label{eq:RNTheta2}
  \Theta_{\mathrm{RN}2} & = & \frac{|\beta|}{\eta^2(\beta^2 + 2 } \bigg[ \frac{8}{3} D^2 \beta^6  + \bigg( \frac{104}{3} D - 16 \bigg) D \beta^4 + \bigg( \frac{272}{3} D^2  \nonumber\\
  & & - 64 D + 32 \bigg) \beta^2 - \frac{64}{3} D^2- \frac{675}{128} \pi^2 + 128 \nonumber\\
  & & -q \bigg( 8\beta^2 - \frac{135}{64}\pi^2 + 32 \bigg) - \frac{27}{128}q\pi^2 \bigg]  .
\end{eqnarray}

The function $T_{\mathrm{RN}}(R)$ in the time delay is
\begin{equation}
  \label{eq:TRN(R)exp}
  T_{\mathrm{RN}}(R)  = T_0 + \sum_{k=1}^{3}\frac{m_{\bullet}^k}{r^k_0} r_0 T_{\mathrm{RN}k} + \mathcal{O}\bigg(\frac{m_{\bullet}^4}{r_0^4}\bigg),
\end{equation}
where
\begin{eqnarray}
  \label{}
  T_{\mathrm{RN}0} & = & \sqrt{R^2-r_0^2},\\
T_{\mathrm{RN}1} & = & \frac{\sqrt{1-\xi^2}}{1+\xi} + 2 \ln\bigg(\frac{1+\sqrt{1-\xi^2}}{\xi}\bigg) ,\\
T_{\mathrm{RN}2} & = & \frac{3}{2}(5-q)\bigg(\frac{\pi}{2}-\arcsin\xi\bigg) - \bigg(2+\frac{5}{2}\xi\bigg)\frac{\sqrt{1-\xi^2}}{(\xi+1)^2},\\
T_{\mathrm{RN}3} & = & -\frac{3}{2}(5-q)\bigg(\frac{\pi}{2}-\arcsin\xi\bigg) \nonumber\\
  & & + \frac{\sqrt{1-\xi^2}}{2(\xi+1)^2} [ 5(7-3q)\xi^3 + (133-52q)\xi^2 \nonumber\\
  & & + (157-59 q) \xi + 60 -22q ].
\end{eqnarray}
It leads to the scaled time delay as
\begin{equation}
  \label{}
  \hat{\tau}_{\mathrm{RN}} = \hat{\tau}_{\mathrm{RN}0} + \varepsilon \hat{\tau}_{\mathrm{RN}1} +\mathcal{O}(\varepsilon^2),
\end{equation}
where
\begin{eqnarray}
  \label{}
  \hat{\tau}_{\mathrm{RN}0} & = & \frac{1}{2} \bigg[1 + \beta^2 - \theta_0^2 - \ln \bigg(\frac{d_{\mathrm{L}}\theta_0^2\vartheta_{\mathrm{E}}^2}{4d_{\mathrm{LS}}}\bigg)\bigg],\\
  \hat{\tau}_{\mathrm{RN}1} & = & \frac{3(5-q)\pi}{16\theta_0} .
\end{eqnarray}
Therefore, the differential time delay between the two lensed images is
\begin{equation}
  \label{}
  \Delta\hat{\tau}_{\mathrm{RN}} = \Delta\hat{\tau}_{\mathrm{RN}0} + \varepsilon \Delta\hat{\tau}_{\mathrm{RN}1} + \mathcal{O}(\varepsilon^2),
\end{equation}
where
\begin{eqnarray}
  \label{}
  \Delta\hat{\tau}_{\mathrm{RN}0} & = &   \frac{1}{2} \eta |\beta| + \ln \bigg( \frac{\eta + |\beta|}{\eta - |\beta|} \bigg), \\
  \Delta\hat{\tau}_{\mathrm{RN}1} & = & \frac{3(5-q)}{16} \pi |\beta| 
\end{eqnarray}

The practical observables for the weak deflection lensing by the \rnbh\ are
\begin{eqnarray}
  \label{eq:RNPtot}
  P_{\mathrm{RN,tot}} &  = & \mathcal{E} + \frac{3(5-q)}{16} \varepsilon \pi \vartheta_{\mathrm{E}}    +\mathcal{O}(\varepsilon^2), \\
  \label{}
  \Delta P_{\mathrm{RN}} &  = & |\mathcal{B}| \bigg( 1 - \frac{3(5-q)}{16} \varepsilon \pi \frac{ \vartheta_{\mathrm{E}} }{ \mathcal{E} }  \bigg)  +\mathcal{O}(\varepsilon^2),\\
  \label{}
  F_{\mathrm{RN,tot}} &  = &  F_{\mathrm{src}} \frac{ \mathcal{B}^2 + 2 \vartheta_{\mathrm{E}}^2 } {|\mathcal{B}| \mathcal{E} } +\mathcal{O}(\varepsilon^2),\\
  \label{}
  \Delta F_{\mathrm{RN}} &  = & F_{\mathrm{src}}- F_{\mathrm{src}}\frac{3(5-q)}{8} \varepsilon \pi \frac{\vartheta_{\mathrm{E}}^3}{\mathcal{E}^{3}}  +\mathcal{O}(\varepsilon^2),\\
  \label{}
  S_{\mathrm{RN,cent}} &  = &  |\mathcal{B}| \frac{\mathcal{B}^2+3 \vartheta_{\mathrm{E}}^2}{\mathcal{B}^2+2 \vartheta_{\mathrm{E}}^2}  +\mathcal{O}(\varepsilon^2),\\
  \label{}
  \Delta \tau & = & \frac{d_{\mathrm{L}}d_{\mathrm{S}}}{cd_{\mathrm{LS}}} \bigg\{ \frac{1}{2} |\mathcal{B}| \mathcal{E}  +\vartheta_{\mathrm{E}}^2 \ln\Bigg( \frac{\mathcal{E}+|\mathcal{B}|}{\mathcal{E}-|\mathcal{B}|} \Bigg) + \varepsilon \frac{3(5-q)}{16} \pi \vartheta_{\mathrm{E}} |\mathcal{B}|   +\mathcal{O}(\varepsilon^2)\bigg\},
\end{eqnarray}
and their deviations from those of the \schbh in GR are
\begin{eqnarray}
  \label{eq:RNdPtot}
  \delta P_{\mathrm{RN,tot}} & = & - \frac{3}{16}q \varepsilon \pi \vartheta_{\mathrm{E}}    +\mathcal{O}(\varepsilon^2),\\
  \label{eq:RNdDP}
  \delta \Delta P_{\mathrm{RN}} &  = &  \frac{3}{16}q \varepsilon \pi |\mathcal{B}| \frac{ \vartheta_{\mathrm{E}} }{ \mathcal{E} }  +\mathcal{O}(\varepsilon^2),\\
  \label{eq:RNdrtot}
  \delta r_{\mathrm{RN,tot}} & = &  \mathcal{O}(\varepsilon^2),\\
  \label{eq:RNdDr}
  \delta \Delta r &  = & \frac{15}{16\ln10}q\varepsilon \pi \frac{\vartheta_{\mathrm{E}}^3}{\mathcal{E}^{3}} +\mathcal{O}(\varepsilon^2),\\
  \label{eq:RNdScent}
  \delta S_{\mathrm{cent}} &  = &  \mathcal{O}(\varepsilon^2),\\
  \label{eq:RNdDtau}
    \delta \Delta \tau & = &  -\frac{3d_{\mathrm{L}}d_{\mathrm{S}}}{16cd_{\mathrm{LS}}}  \varepsilon q \pi \vartheta_{\mathrm{E}} |\mathcal{B}|   +\mathcal{O}(\varepsilon^2).
\end{eqnarray}

The tidal \rnbh\ shares the same formulae in this appendix but with a negative $q$.

\acknowledgments

This work is funded by the National Natural Science Foundation of China (Grant Nos. 11573015 and 11833004).

\bibliographystyle{JHEP}
\bibliography{Refs20190514}

\providecommand{\href}[2]{#2}\begingroup\raggedright\begin{thebibliography}{10%
0}

\bibitem{Horndeski1974IJTP10.363}
G.~W. {Horndeski}, \emph{{Second-Order Scalar-Tensor Field Equations in a
  Four-Dimensional Space}},
  \href{https://doi.org/10.1007/BF01807638}{\emph{\ijtp} {\bfseries 10} (1974)
  363}.

\bibitem{Lovelock1971JMP12.498}
D.~{Lovelock}, \emph{{The Einstein Tensor and Its Generalizations}},
  \href{https://doi.org/10.1063/1.1665613}{\emph{\jmp} {\bfseries 12} (1971)
  498}.

\bibitem{Rham2018PRL121.221101}
C.~{de Rham} and S.~{Melville}, \emph{{Gravitational Rainbows: LIGO and Dark
  Energy at its Cutoff}},
  \href{https://doi.org/10.1103/PhysRevLett.121.221101}{\emph{\prl} {\bfseries
  121} (2018) 221101} [\href{https://arxiv.org/abs/1806.09417}{{\ttfamily
  1806.09417}}].

\bibitem{Riess1998AJ116.1009}
A.~G. {Riess}, A.~V. {Filippenko}, P.~{Challis}, A.~{Clocchiatti},
  A.~{Diercks}, P.~M. {Garnavich} et~al., \emph{{Observational Evidence from
  Supernovae for an Accelerating Universe and a Cosmological Constant}},
  \href{https://doi.org/10.1086/300499}{\emph{\aj} {\bfseries 116} (1998)
  1009}.

\bibitem{Perlmutter1999ApJ517.565}
{\scshape Supernova Cosmology Project} collaboration, \emph{{Measurements of
  Omega and Lambda from 42 High-Redshift Supernovae}},
  \href{https://doi.org/10.1086/307221}{\emph{\apj} {\bfseries 517} (1999)
  565}.

\bibitem{Clifton2012PhR513.1}
T.~{Clifton}, P.~G. {Ferreira}, A.~{Padilla} and C.~{Skordis}, \emph{{Modified
  gravity and cosmology}},
  \href{https://doi.org/10.1016/j.physrep.2012.01.001}{\emph{\physrep}
  {\bfseries 513} (2012) 1}.

\bibitem{Brans1961PR124.925}
C.~{Brans} and R.~H. {Dicke}, \emph{{Mach's Principle and a Relativistic Theory
  of Gravitation}}, \href{https://doi.org/10.1103/PhysRev.124.925}{\emph{\pr}
  {\bfseries 124} (1961) 925}.

\bibitem{Felice2010LRR13.3}
A.~{De Felice} and S.~{Tsujikawa}, \emph{{f(R) Theories}}, {\emph{\lrr}
  {\bfseries 13} (2010) 3} [\href{https://arxiv.org/abs/1002.4928}{{\ttfamily
  1002.4928}}].

\bibitem{Sotiriou2010RMP82.451}
T.~P. {Sotiriou} and V.~{Faraoni}, \emph{{f(R) theories of gravity}},
  \href{https://doi.org/10.1103/RevModPhys.82.451}{\emph{\rmp} {\bfseries 82}
  (2010) 451} [\href{https://arxiv.org/abs/0805.1726}{{\ttfamily 0805.1726}}].

\bibitem{Dvali2000PLB485.208}
G.~{Dvali}, G.~{Gabadadze} and M.~{Porrati}, \emph{{4D gravity on a brane in 5D
  Minkowski space}},
  \href{https://doi.org/10.1016/S0370-2693(00)00669-9}{\emph{\plb} {\bfseries
  485} (2000) 208} [\href{https://arxiv.org/abs/hep-th/0005016}{{\ttfamily
  hep-th/0005016}}].

\bibitem{Deffayet2009PRD79.084003}
C.~{Deffayet}, G.~{Esposito-Far{\`e}se} and A.~{Vikman}, \emph{{Covariant
  Galileon}}, \href{https://doi.org/10.1103/PhysRevD.79.084003}{\emph{\prd}
  {\bfseries 79} (2009) 084003}
  [\href{https://arxiv.org/abs/0901.1314}{{\ttfamily 0901.1314}}].

\bibitem{Deffayet2009PRD80.064015}
C.~{Deffayet}, S.~{Deser} and G.~{Esposito-Far{\`e}se}, \emph{{Generalized
  Galileons: All scalar models whose curved background extensions maintain
  second-order field equations and stress tensors}},
  \href{https://doi.org/10.1103/PhysRevD.80.064015}{\emph{\prd} {\bfseries 80}
  (2009) 064015} [\href{https://arxiv.org/abs/0906.1967}{{\ttfamily
  0906.1967}}].

\bibitem{Kobayashi2011PTP126.511}
T.~{Kobayashi}, M.~{Yamaguchi} and J.~{Yokoyama}, \emph{{Generalized
  G-Inflation -- Inflation with the Most General Second-Order Field
  Equations}}, \href{https://doi.org/10.1143/PTP.126.511}{\emph{\ptp}
  {\bfseries 126} (2011) 511}
  [\href{https://arxiv.org/abs/1105.5723}{{\ttfamily 1105.5723}}].

\bibitem{Deffayet2013CQG30.214006}
C.~{Deffayet} and D.~A. {Steer}, \emph{{A formal introduction to Horndeski and
  Galileon theories and their generalizations}},
  \href{https://doi.org/10.1088/0264-9381/30/21/214006}{\emph{\cqg} {\bfseries
  30} (2013) 214006} [\href{https://arxiv.org/abs/1307.2450}{{\ttfamily
  1307.2450}}].

\bibitem{Abbott2017PRL119.161101}
B.~P. {Abbott}, R.~{Abbott}, T.~D. {Abbott}, F.~{Acernese}, K.~{Ackley},
  C.~{Adams} et~al., \emph{{GW170817: Observation of Gravitational Waves from a
  Binary Neutron Star Inspiral}},
  \href{https://doi.org/10.1103/PhysRevLett.119.161101}{\emph{\prl} {\bfseries
  119} (2017) 161101} [\href{https://arxiv.org/abs/1710.05832}{{\ttfamily
  1710.05832}}].

\bibitem{Abbott2017ApJ848.L12}
B.~P. {Abbott}, R.~{Abbott}, T.~D. {Abbott}, F.~{Acernese}, K.~{Ackley},
  C.~{Adams} et~al., \emph{{Multi-messenger Observations of a Binary Neutron
  Star Merger}}, \href{https://doi.org/10.3847/2041-8213/aa91c9}{\emph{\apjl}
  {\bfseries 848} (2017) L12}
  [\href{https://arxiv.org/abs/1710.05833}{{\ttfamily 1710.05833}}].

\bibitem{Abbott2017ApJ848.L13}
B.~P. {Abbott}, R.~{Abbott}, T.~D. {Abbott}, F.~{Acernese}, K.~{Ackley},
  C.~{Adams} et~al., \emph{{Gravitational Waves and Gamma-Rays from a Binary
  Neutron Star Merger: GW170817 and GRB 170817A}},
  \href{https://doi.org/10.3847/2041-8213/aa920c}{\emph{\apjl} {\bfseries 848}
  (2017) L13} [\href{https://arxiv.org/abs/1710.05834}{{\ttfamily
  1710.05834}}].

\bibitem{Lombriser2017PLB765.382}
L.~{Lombriser} and N.~A. {Lima}, \emph{{Challenges to self-acceleration in
  modified gravity from gravitational waves and large-scale structure}},
  \href{https://doi.org/10.1016/j.physletb.2016.12.048}{\emph{\plb} {\bfseries
  765} (2017) 382} [\href{https://arxiv.org/abs/1602.07670}{{\ttfamily
  1602.07670}}].

\bibitem{Lombriser2016JCAP03.031}
L.~{Lombriser} and A.~{Taylor}, \emph{{Breaking a dark degeneracy with
  gravitational waves}},
  \href{https://doi.org/10.1088/1475-7516/2016/03/031}{\emph{\jcap} {\bfseries
  03} (2016) 031} [\href{https://arxiv.org/abs/1509.08458}{{\ttfamily
  1509.08458}}].

\bibitem{Baker2017PRL119.251301}
T.~{Baker}, E.~{Bellini}, P.~G. {Ferreira}, M.~{Lagos}, J.~{Noller} and
  I.~{Sawicki}, \emph{{Strong Constraints on Cosmological Gravity from GW170817
  and GRB 170817A}},
  \href{https://doi.org/10.1103/PhysRevLett.119.251301}{\emph{\prl} {\bfseries
  119} (2017) 251301} [\href{https://arxiv.org/abs/1710.06394}{{\ttfamily
  1710.06394}}].

\bibitem{Creminelli2017PRL119.251302}
P.~{Creminelli} and F.~{Vernizzi}, \emph{{Dark Energy after GW170817 and
  GRB170817A}},
  \href{https://doi.org/10.1103/PhysRevLett.119.251302}{\emph{\prl} {\bfseries
  119} (2017) 251302} [\href{https://arxiv.org/abs/1710.05877}{{\ttfamily
  1710.05877}}].

\bibitem{Sakstein2017PRL119.251303}
J.~{Sakstein} and B.~{Jain}, \emph{{Implications of the Neutron Star Merger
  GW170817 for Cosmological Scalar-Tensor Theories}},
  \href{https://doi.org/10.1103/PhysRevLett.119.251303}{\emph{\prl} {\bfseries
  119} (2017) 251303} [\href{https://arxiv.org/abs/1710.05893}{{\ttfamily
  1710.05893}}].

\bibitem{Ezquiaga2017PRL119.251304}
J.~M. {Ezquiaga} and M.~{Zumalac{\'a}rregui}, \emph{{Dark Energy After
  GW170817: Dead Ends and the Road Ahead}},
  \href{https://doi.org/10.1103/PhysRevLett.119.251304}{\emph{\prl} {\bfseries
  119} (2017) 251304} [\href{https://arxiv.org/abs/1710.05901}{{\ttfamily
  1710.05901}}].

\bibitem{Casalino2018PDU22.108}
A.~{Casalino}, M.~{Rinaldi}, L.~{Sebastiani} and S.~{Vagnozzi},
  \emph{{Mimicking dark matter and dark energy in a mimetic model compatible
  with GW170817}},
  \href{https://doi.org/10.1016/j.dark.2018.10.001}{\emph{\pdu} {\bfseries 22}
  (2018) 108} [\href{https://arxiv.org/abs/1803.02620}{{\ttfamily
  1803.02620}}].

\bibitem{Casalino2019CQG36.017001}
A.~{Casalino}, M.~{Rinaldi}, L.~{Sebastiani} and S.~{Vagnozzi}, \emph{{Alive
  and well: mimetic gravity and a higher-order extension in light of
  GW170817}}, \href{https://doi.org/10.1088/1361-6382/aaf1fd}{\emph{\cqg}
  {\bfseries 36} (2019) 017001}
  [\href{https://arxiv.org/abs/1811.06830}{{\ttfamily 1811.06830}}].

\bibitem{Ishak2019LRR22.1}
M.~{Ishak}, \emph{{Testing general relativity in cosmology}},
  \href{https://doi.org/10.1007/s41114-018-0017-4}{\emph{\lrr} {\bfseries 22}
  (2019) 1} [\href{https://arxiv.org/abs/1806.10122}{{\ttfamily 1806.10122}}].

\bibitem{Hohmann2015PRD92.064019}
M.~{Hohmann}, \emph{{Parametrized post-Newtonian limit of Horndeski's gravity
  theory}}, \href{https://doi.org/10.1103/PhysRevD.92.064019}{\emph{\prd}
  {\bfseries 92} (2015) 064019}
  [\href{https://arxiv.org/abs/1506.04253}{{\ttfamily 1506.04253}}].

\bibitem{Bhattacharya2017PRD95.044037}
S.~{Bhattacharya} and S.~{Chakraborty}, \emph{{Constraining some Horndeski
  gravity theories}},
  \href{https://doi.org/10.1103/PhysRevD.95.044037}{\emph{\prd} {\bfseries 95}
  (2017) 044037} [\href{https://arxiv.org/abs/1607.03693}{{\ttfamily
  1607.03693}}].

\bibitem{Hou2018EPJC78.247}
S.~{Hou} and Y.~{Gong}, \emph{{Constraints on Horndeski theory using the
  observations of Nordtvedt effect, Shapiro time delay and binary pulsars}},
  \href{https://doi.org/10.1140/epjc/s10052-018-5738-8}{\emph{\epjc} {\bfseries
  78} (2018) 247} [\href{https://arxiv.org/abs/1711.05034}{{\ttfamily
  1711.05034}}].

\bibitem{Dyadina2019MNRAS483.947}
P.~I. {Dyadina}, N.~A. {Avdeev} and S.~O. {Alexeyev}, \emph{{Horndeski gravity
  without screening in binary pulsars}},
  \href{https://doi.org/10.1093/mnras/sty3094}{\emph{\mnras} {\bfseries 483}
  (2019) 947} [\href{https://arxiv.org/abs/1811.05393}{{\ttfamily
  1811.05393}}].

\bibitem{Cisterna2015PRD92.044050}
A.~{Cisterna}, T.~{Delsate} and M.~{Rinaldi}, \emph{{Neutron stars in general
  second order scalar-tensor theory: The case of nonminimal derivative
  coupling}}, \href{https://doi.org/10.1103/PhysRevD.92.044050}{\emph{\prd}
  {\bfseries 92} (2015) 044050}
  [\href{https://arxiv.org/abs/1504.05189}{{\ttfamily 1504.05189}}].

\bibitem{Cisterna2016PRD93.084046}
A.~{Cisterna}, T.~{Delsate}, L.~{Ducobu} and M.~{Rinaldi}, \emph{{Slowly
  rotating neutron stars in the nonminimal derivative coupling sector of
  Horndeski gravity}},
  \href{https://doi.org/10.1103/PhysRevD.93.084046}{\emph{\prd} {\bfseries 93}
  (2016) 084046} [\href{https://arxiv.org/abs/1602.06939}{{\ttfamily
  1602.06939}}].

\bibitem{Babichev2016CQG33.154002}
E.~{Babichev}, C.~{Charmousis} and A.~{Leh{\'e}bel}, \emph{{Black holes and
  stars in Horndeski theory}},
  \href{https://doi.org/10.1088/0264-9381/33/15/154002}{\emph{\cqg} {\bfseries
  33} (2016) 154002} [\href{https://arxiv.org/abs/1604.06402}{{\ttfamily
  1604.06402}}].

\bibitem{Maselli2016PRD93.124056}
A.~{Maselli}, H.~O. {Silva}, M.~{Minamitsuji} and E.~{Berti}, \emph{{Neutron
  stars in Horndeski gravity}},
  \href{https://doi.org/10.1103/PhysRevD.93.124056}{\emph{\prd} {\bfseries 93}
  (2016) 124056} [\href{https://arxiv.org/abs/1603.04876}{{\ttfamily
  1603.04876}}].

\bibitem{Silva2016IJMPD25.1641006}
H.~O. {Silva}, A.~{Maselli}, M.~{Minamitsuji} and E.~{Berti}, \emph{{Compact
  objects in Horndeski gravity}},
  \href{https://doi.org/10.1142/S0218271816410066}{\emph{\ijmpd} {\bfseries 25}
  (2016) 1641006} [\href{https://arxiv.org/abs/1602.05997}{{\ttfamily
  1602.05997}}].

\bibitem{Lehebel2017JCAP07.037}
A.~{Leh{\'e}bel}, E.~{Babichev} and C.~{Charmousis}, \emph{{A no-hair theorem
  for stars in Horndeski theories}},
  \href{https://doi.org/10.1088/1475-7516/2017/07/037}{\emph{\jcap} {\bfseries
  07} (2017) 037} [\href{https://arxiv.org/abs/1706.04989}{{\ttfamily
  1706.04989}}].

\bibitem{Rinaldi2012PRD86.084048}
M.~{Rinaldi}, \emph{{Black holes with nonminimal derivative coupling}},
  \href{https://doi.org/10.1103/PhysRevD.86.084048}{\emph{\prd} {\bfseries 86}
  (2012) 084048} [\href{https://arxiv.org/abs/1208.0103}{{\ttfamily
  1208.0103}}].

\bibitem{Anabalon2014PRD89.084050}
A.~{Anabalon}, A.~{Cisterna} and J.~{Oliva}, \emph{{Asymptotically locally AdS
  and flat black holes in Horndeski theory}},
  \href{https://doi.org/10.1103/PhysRevD.89.084050}{\emph{\prd} {\bfseries 89}
  (2014) 084050} [\href{https://arxiv.org/abs/1312.3597}{{\ttfamily
  1312.3597}}].

\bibitem{Cisterna2014PRD89.084038}
A.~{Cisterna} and C.~{Erices}, \emph{{Asymptotically locally AdS and flat black
  holes in the presence of an electric field in the Horndeski scenario}},
  \href{https://doi.org/10.1103/PhysRevD.89.084038}{\emph{\prd} {\bfseries 89}
  (2014) 084038} [\href{https://arxiv.org/abs/1401.4479}{{\ttfamily
  1401.4479}}].

\bibitem{Babichev2015JCAP5.31}
E.~{Babichev}, C.~{Charmousis} and M.~{Hassaine}, \emph{{Charged Galileon black
  holes}}, \href{https://doi.org/10.1088/1475-7516/2015/05/031}{\emph{\jcap}
  {\bfseries 05} (2015) 31} [\href{https://arxiv.org/abs/1503.02545}{{\ttfamily
  1503.02545}}].

\bibitem{Babichev2016JCAP09.011}
E.~{Babichev}, C.~{Charmousis}, A.~{Leh{\'e}bel} and T.~{Moskalets},
  \emph{{Black holes in a cubic Galileon universe}},
  \href{https://doi.org/10.1088/1475-7516/2016/09/011}{\emph{\jcap} {\bfseries
  09} (2016) 011} [\href{https://arxiv.org/abs/1605.07438}{{\ttfamily
  1605.07438}}].

\bibitem{Babichev2017JCAP04.027}
E.~{Babichev}, C.~{Charmousis} and A.~{Leh{\'e}bel}, \emph{{Asymptotically flat
  black holes in Horndeski theory and beyond}},
  \href{https://doi.org/10.1088/1475-7516/2017/04/027}{\emph{\jcap} {\bfseries
  04} (2017) 027} [\href{https://arxiv.org/abs/1702.01938}{{\ttfamily
  1702.01938}}].

\bibitem{Antoniou2018PRD97.084037}
G.~{Antoniou}, A.~{Bakopoulos} and P.~{Kanti}, \emph{{Black-hole solutions with
  scalar hair in Einstein-scalar-Gauss-Bonnet theories}},
  \href{https://doi.org/10.1103/PhysRevD.97.084037}{\emph{\prd} {\bfseries 97}
  (2018) 084037} [\href{https://arxiv.org/abs/1711.07431}{{\ttfamily
  1711.07431}}].

\bibitem{Bakopoulos2019PRD99.064003}
A.~Bakopoulos, G.~Antoniou and P.~Kanti, \emph{Novel black-hole solutions in
  einstein-scalar-gauss-bonnet theories with a cosmological constant},
  \href{https://doi.org/10.1103/PhysRevD.99.064003}{\emph{\prd} {\bfseries 99}
  (2019) 064003}.

\bibitem{Hui2013PRL110.241104}
L.~{Hui} and A.~{Nicolis}, \emph{{No-Hair Theorem for the Galileon}},
  \href{https://doi.org/10.1103/PhysRevLett.110.241104}{\emph{\prl} {\bfseries
  110} (2013) 241104} [\href{https://arxiv.org/abs/1202.1296}{{\ttfamily
  1202.1296}}].

\bibitem{Sotiriou2014PRL112.251102}
T.~P. {Sotiriou} and S.-Y. {Zhou}, \emph{{Black Hole Hair in Generalized
  Scalar-Tensor Gravity}},
  \href{https://doi.org/10.1103/PhysRevLett.112.251102}{\emph{\prl} {\bfseries
  112} (2014) 251102} [\href{https://arxiv.org/abs/1312.3622}{{\ttfamily
  1312.3622}}].

\bibitem{Sotiriou2014PRD90.124063}
T.~P. {Sotiriou} and S.-Y. {Zhou}, \emph{{Black hole hair in generalized
  scalar-tensor gravity: An explicit example}},
  \href{https://doi.org/10.1103/PhysRevD.90.124063}{\emph{\prd} {\bfseries 90}
  (2014) 124063} [\href{https://arxiv.org/abs/1408.1698}{{\ttfamily
  1408.1698}}].

\bibitem{Antoniou2018PRL120.131102}
G.~{Antoniou}, A.~{Bakopoulos} and P.~{Kanti}, \emph{{Evasion of No-Hair
  Theorems and Novel Black-Hole Solutions in Gauss-Bonnet Theories}},
  \href{https://doi.org/10.1103/PhysRevLett.120.131102}{\emph{\prl} {\bfseries
  120} (2018) 131102} [\href{https://arxiv.org/abs/1711.03390}{{\ttfamily
  1711.03390}}].

\bibitem{Anabalon2014PRD90.124055}
A.~{Anabal{\'o}n}, J.~{Bi{\v c}{\'a}k} and J.~{Saavedra}, \emph{{Hairy black
  holes: Stability under odd-parity perturbations and existence of slowly
  rotating solutions}},
  \href{https://doi.org/10.1103/PhysRevD.90.124055}{\emph{\prd} {\bfseries 90}
  (2014) 124055} [\href{https://arxiv.org/abs/1405.7893}{{\ttfamily
  1405.7893}}].

\bibitem{Cisterna2015PRD92.104018}
A.~{Cisterna}, M.~{Cruz}, T.~{Delsate} and J.~{Saavedra}, \emph{{Nonminimal
  derivative coupling scalar-tensor theories: Odd-parity perturbations and
  black hole stability}},
  \href{https://doi.org/10.1103/PhysRevD.92.104018}{\emph{\prd} {\bfseries 92}
  (2015) 104018} [\href{https://arxiv.org/abs/1508.06413}{{\ttfamily
  1508.06413}}].

\bibitem{Miao2016EPJC76.638}
Y.-G. {Miao} and Z.-M. {Xu}, \emph{{Thermodynamics of Horndeski black holes
  with non-minimal derivative coupling}},
  \href{https://doi.org/10.1140/epjc/s10052-016-4482-1}{\emph{\epjc} {\bfseries
  76} (2016) 638} [\href{https://arxiv.org/abs/1607.06629}{{\ttfamily
  1607.06629}}].

\bibitem{Feng2016PRD93.044030}
X.-H. {Feng}, H.-S. {Liu}, H.~{L{\"u}} and C.~N. {Pope}, \emph{{Thermodynamics
  of charged black holes in Einstein-Horndeski-Maxwell theory}},
  \href{https://doi.org/10.1103/PhysRevD.93.044030}{\emph{\prd} {\bfseries 93}
  (2016) 044030} [\href{https://arxiv.org/abs/1512.02659}{{\ttfamily
  1512.02659}}].

\bibitem{Perlick2004LRR7.9}
V.~{Perlick}, \emph{{Gravitational Lensing from a Spacetime Perspective}},
  \href{https://doi.org/10.12942/lrr-2004-9}{\emph{\lrr} {\bfseries 7} (2004)
  9}.

\bibitem{Schneider1992GL}
P.~{Schneider}, J.~{Ehlers} and E.~E. {Falco}, \emph{{Gravitational Lenses}}.
  Springer-Verlag, Berlin, 1992,
  \href{https://doi.org/10.1007/978-3-662-03758-4}{10.1007/978-3-662-03758-4}.

\bibitem{Petters2001STGL}
A.~O. {Petters}, H.~{Levine} and J.~{Wambsganss}, \emph{{Singularity Theory and
  Gravitational Lensing}}. Birkh{\"a}user, Boston, 2001,
  \href{https://doi.org/10.1007/978-1-4612-0145-8}{10.1007/978-1-4612-0145-8}.

\bibitem{Schneider2006GLSWM}
P.~{Schneider}, C.~S. {Kochanek} and J.~{Wambsganss}, \emph{{Gravitational
  Lensing: Strong, Weak and Micro}},  in \emph{Saas-Fee Advanced Course 33:
  Gravitational Lensing: Strong, Weak and Micro}, G.~{Meylan}, P.~{Jetzer} and
  P.~{North}, eds., (Berlin), Springer-Verlag, 2006,
  \href{https://doi.org/10.1007/978-3-540-30310-7}{DOI}.

\bibitem{Aubourg1993Nature365.623}
E.~{Aubourg}, P.~{Bareyre}, S.~{Br{\'e}hin}, M.~{Gros}, M.~{Lachi{\`e}ze-Rey},
  B.~{Laurent} et~al., \emph{{Evidence for gravitational microlensing by dark
  objects in the Galactic halo}},
  \href{https://doi.org/10.1038/365623a0}{\emph{\nat} {\bfseries 365} (1993)
  623}.

\bibitem{Sahu2017Sci356.1046}
K.~C. {Sahu}, J.~{Anderson}, S.~{Casertano}, H.~E. {Bond}, P.~{Bergeron}, E.~P.
  {Nelan} et~al., \emph{{Relativistic deflection of background starlight
  measures the mass of a nearby white dwarf star}},
  \href{https://doi.org/10.1126/science.aal2879}{\emph{\sci} {\bfseries 356}
  (2017) 1046} [\href{https://arxiv.org/abs/1706.02037}{{\ttfamily
  1706.02037}}].

\bibitem{Keeton2005PRD72.104006}
C.~R. {Keeton} and A.~O. {Petters}, \emph{{Formalism for testing theories of
  gravity using lensing by compact objects: Static, spherically symmetric
  case}}, \href{https://doi.org/10.1103/PhysRevD.72.104006}{\emph{\prd}
  {\bfseries 72} (2005) 104006}
  [\href{https://arxiv.org/abs/gr-qc/0511019}{{\ttfamily gr-qc/0511019}}].

\bibitem{Keeton2006PRD73.044024}
C.~R. {Keeton} and A.~O. {Petters}, \emph{{Formalism for testing theories of
  gravity using lensing by compact objects. II. Probing post-post-Newtonian
  metrics}}, \href{https://doi.org/10.1103/PhysRevD.73.044024}{\emph{\prd}
  {\bfseries 73} (2006) 044024}
  [\href{https://arxiv.org/abs/gr-qc/0601053}{{\ttfamily gr-qc/0601053}}].

\bibitem{Keeton2006PRD73.104032}
C.~R. {Keeton} and A.~O. {Petters}, \emph{{Formalism for testing theories of
  gravity using lensing by compact objects. III. Braneworld gravity}},
  \href{https://doi.org/10.1103/PhysRevD.73.104032}{\emph{\prd} {\bfseries 73}
  (2006) 104032} [\href{https://arxiv.org/abs/gr-qc/0603061}{{\ttfamily
  gr-qc/0603061}}].

\bibitem{Collett2018Sci360.1342}
T.~E. {Collett}, L.~J. {Oldham}, R.~J. {Smith}, M.~W. {Auger}, K.~B.
  {Westfall}, D.~{Bacon} et~al., \emph{{A precise extragalactic test of General
  Relativity}}, \href{https://doi.org/10.1126/science.aao2469}{\emph{\sci}
  {\bfseries 360} (2018) 1342}
  [\href{https://arxiv.org/abs/1806.08300}{{\ttfamily 1806.08300}}].

\bibitem{Li2017AoP382.136}
G.~{Li} and X.-M. {Deng}, \emph{{Classical tests of photons coupled to Weyl
  tensor in the Solar System}},
  \href{https://doi.org/10.1016/j.aop.2017.05.001}{\emph{\aop} {\bfseries 382}
  (2017) 136}.

\bibitem{Cao2018EPJC78.191}
W.-G. {Cao} and Y.~{Xie}, \emph{{Weak deflection gravitational lensing for
  photons coupled to Weyl tensor in a Schwarzschild black hole}},
  \href{https://doi.org/10.1140/epjc/s10052-018-5684-5}{\emph{\epjc} {\bfseries
  78} (2018) 191}.

\bibitem{Bozza2010GRG42.2269}
V.~{Bozza}, \emph{{Gravitational lensing by black holes}},
  \href{https://doi.org/10.1007/s10714-010-0988-2}{\emph{\grg} {\bfseries 42}
  (2010) 2269} [\href{https://arxiv.org/abs/0911.2187}{{\ttfamily 0911.2187}}].

\bibitem{Eiroa2012arXiv1212.4535}
E.~F. {Eiroa}, \emph{{Strong deflection gravitational lensing}},
  {\emph{arXiv:1212.4535} (2012) }
  [\href{https://arxiv.org/abs/1212.4535}{{\ttfamily 1212.4535}}].

\bibitem{Cunha2018GRG50.42}
P.~V.~P. {Cunha} and C.~A.~R. {Herdeiro}, \emph{{Shadows and strong
  gravitational lensing: a brief review}},
  \href{https://doi.org/10.1007/s10714-018-2361-9}{\emph{\grg} {\bfseries 50}
  (2018) 42} [\href{https://arxiv.org/abs/1801.00860}{{\ttfamily 1801.00860}}].

\bibitem{Darwin1959PRSLSA249.180}
C.~{Darwin}, \emph{{The Gravity Field of a Particle}},
  \href{https://doi.org/10.1098/rspa.1959.0015}{\emph{Proc. R. Soc. Lond. Ser.
  A} {\bfseries 249} (1959) 180}.

\bibitem{Luminet1979AA75.228}
J.-P. {Luminet}, \emph{{Image of a spherical black hole with thin accretion
  disk}}, {\emph{\aap} {\bfseries 75} (1979) 228}.

\bibitem{Ohanian1987AJP55.428}
H.~C. {Ohanian}, \emph{{The black hole as a gravitational ``lens''}},
  \href{https://doi.org/10.1119/1.15126}{\emph{\ajp} {\bfseries 55} (1987)
  428}.

\bibitem{Nemiroff1993AJP61.619}
R.~J. {Nemiroff}, \emph{{Visual distortions near a neutron star and black
  hole}}, \href{https://doi.org/10.1119/1.17224}{\emph{\ajp} {\bfseries 61}
  (1993) 619} [\href{https://arxiv.org/abs/astro-ph/9312003}{{\ttfamily
  astro-ph/9312003}}].

\bibitem{Bozza2001GRG33.1535}
V.~{Bozza}, S.~{Capozziello}, G.~{Iovane} and G.~{Scarpetta}, \emph{{Strong
  Field Limit of Black Hole Gravitational Lensing}},
  \href{https://doi.org/10.1023/A:1012292927358}{\emph{\grg} {\bfseries 33}
  (2001) 1535} [\href{https://arxiv.org/abs/gr-qc/0102068}{{\ttfamily
  gr-qc/0102068}}].

\bibitem{Virbhadra1998AA337.1}
K.~S. {Virbhadra}, D.~{Narasimha} and S.~M. {Chitre}, \emph{{Role of the scalar
  field in gravitational lensing}}, {\emph{\aap} {\bfseries 337} (1998) 1}
  [\href{https://arxiv.org/abs/astro-ph/9801174}{{\ttfamily
  astro-ph/9801174}}].

\bibitem{Virbhadra2000PRD62.084003}
K.~S. {Virbhadra} and G.~F.~R. {Ellis}, \emph{{Schwarzschild black hole
  lensing}}, {\emph{\prd} {\bfseries 62} (2000) 084003}
  [\href{https://arxiv.org/abs/astro-ph/9904193}{{\ttfamily
  astro-ph/9904193}}].

\bibitem{Bozza2002PRD66.103001}
V.~{Bozza}, \emph{{Gravitational lensing in the strong field limit}},
  \href{https://doi.org/10.1103/PhysRevD.66.103001}{\emph{\prd} {\bfseries 66}
  (2002) 103001} [\href{https://arxiv.org/abs/gr-qc/0208075}{{\ttfamily
  gr-qc/0208075}}].

\bibitem{Eiroa2002PRD66.024010}
E.~F. {Eiroa}, G.~E. {Romero} and D.~F. {Torres}, \emph{{Reissner-Nordstr{\"o}m
  black hole lensing}},
  \href{https://doi.org/10.1103/PhysRevD.66.024010}{\emph{\prd} {\bfseries 66}
  (2002) 024010} [\href{https://arxiv.org/abs/gr-qc/0203049}{{\ttfamily
  gr-qc/0203049}}].

\bibitem{Bhadra2003PRD67.103009}
A.~{Bhadra}, \emph{{Gravitational lensing by a charged black hole of string
  theory}}, \href{https://doi.org/10.1103/PhysRevD.67.103009}{\emph{\prd}
  {\bfseries 67} (2003) 103009}
  [\href{https://arxiv.org/abs/gr-qc/0306016}{{\ttfamily gr-qc/0306016}}].

\bibitem{Bozza2004GRG36.435}
V.~{Bozza} and L.~{Mancini}, \emph{{Time Delay in Black Hole Gravitational
  Lensing as a Distance Estimator}},
  \href{https://doi.org/10.1023/B:GERG.0000010486.58026.4f}{\emph{\grg}
  {\bfseries 36} (2004) 435}
  [\href{https://arxiv.org/abs/gr-qc/0305007}{{\ttfamily gr-qc/0305007}}].

\bibitem{Perlick2004PRD69.064017}
V.~{Perlick}, \emph{{Exact gravitational lens equation in spherically symmetric
  and static spacetimes}},
  \href{https://doi.org/10.1103/PhysRevD.69.064017}{\emph{\prd} {\bfseries 69}
  (2004) 064017} [\href{https://arxiv.org/abs/gr-qc/0307072}{{\ttfamily
  gr-qc/0307072}}].

\bibitem{Whisker2005PRD71.064004}
R.~{Whisker}, \emph{{Strong gravitational lensing by braneworld black holes}},
  \href{https://doi.org/10.1103/PhysRevD.71.064004}{\emph{\prd} {\bfseries 71}
  (2005) 064004} [\href{https://arxiv.org/abs/astro-ph/0411786}{{\ttfamily
  astro-ph/0411786}}].

\bibitem{Majumdar2005IJMPD14.1095}
A.~S. {Majumdar} and N.~{Mukherjee}, \emph{{Braneworld Black Holes in Cosmology
  and Astrophysics}},
  \href{https://doi.org/10.1142/S0218271805006948}{\emph{\ijmpd} {\bfseries 14}
  (2005) 1095} [\href{https://arxiv.org/abs/astro-ph/0503473}{{\ttfamily
  astro-ph/0503473}}].

\bibitem{Eiroa2004PRD69.063004}
E.~F. {Eiroa} and D.~F. {Torres}, \emph{{Strong field limit analysis of
  gravitational retrolensing}},
  \href{https://doi.org/10.1103/PhysRevD.69.063004}{\emph{\prd} {\bfseries 69}
  (2004) 063004} [\href{https://arxiv.org/abs/gr-qc/0311013}{{\ttfamily
  gr-qc/0311013}}].

\bibitem{Eiroa2005PRD71.083010}
E.~F. {Eiroa}, \emph{{Braneworld black hole gravitational lens: Strong field
  limit analysis}},
  \href{https://doi.org/10.1103/PhysRevD.71.083010}{\emph{\prd} {\bfseries 71}
  (2005) 083010} [\href{https://arxiv.org/abs/gr-qc/0410128}{{\ttfamily
  gr-qc/0410128}}].

\bibitem{Eiroa2006PRD73.043002}
E.~F. {Eiroa}, \emph{{Gravitational lensing by Einstein-Born-Infeld black
  holes}}, \href{https://doi.org/10.1103/PhysRevD.73.043002}{\emph{\prd}
  {\bfseries 73} (2006) 043002}
  [\href{https://arxiv.org/abs/gr-qc/0511065}{{\ttfamily gr-qc/0511065}}].

\bibitem{Iyer2007GRG39.1563}
S.~V. {Iyer} and A.~O. {Petters}, \emph{{Light's bending angle due to black
  holes: from the photon sphere to infinity}},
  \href{https://doi.org/10.1007/s10714-007-0481-8}{\emph{\grg} {\bfseries 39}
  (2007) 1563} [\href{https://arxiv.org/abs/gr-qc/0611086}{{\ttfamily
  gr-qc/0611086}}].

\bibitem{Mukherjee2007GRG39.583}
N.~{Mukherjee} and A.~S. {Majumdar}, \emph{{Particle motion and gravitational
  lensing in the metric of a dilaton black hole in a de Sitter universe}},
  \href{https://doi.org/10.1007/s10714-007-0407-5}{\emph{\grg} {\bfseries 39}
  (2007) 583} [\href{https://arxiv.org/abs/astro-ph/0605224}{{\ttfamily
  astro-ph/0605224}}].

\bibitem{Bozza2008PRD78.103005}
V.~{Bozza}, \emph{{Comparison of approximate gravitational lens equations and a
  proposal for an improved new one}},
  \href{https://doi.org/10.1103/PhysRevD.78.103005}{\emph{\prd} {\bfseries 78}
  (2008) 103005} [\href{https://arxiv.org/abs/0807.3872}{{\ttfamily
  0807.3872}}].

\bibitem{Virbhadra2008PRD77.124014}
K.~S. {Virbhadra} and C.~R. {Keeton}, \emph{{Time delay and magnification
  centroid due to gravitational lensing by black holes and naked
  singularities}},
  \href{https://doi.org/10.1103/PhysRevD.77.124014}{\emph{\prd} {\bfseries 77}
  (2008) 124014} [\href{https://arxiv.org/abs/0710.2333}{{\ttfamily
  0710.2333}}].

\bibitem{Virbhadra2009PRD79.083004}
K.~S. {Virbhadra}, \emph{{Relativistic images of Schwarzschild black hole
  lensing}}, \href{https://doi.org/10.1103/PhysRevD.79.083004}{\emph{\prd}
  {\bfseries 79} (2009) 083004}
  [\href{https://arxiv.org/abs/0810.2109}{{\ttfamily 0810.2109}}].

\bibitem{Bin-Nun2010PRD82.064009}
A.~Y. {Bin-Nun}, \emph{{Gravitational lensing of stars orbiting Sgr A* as a
  probe of the black hole metric in the Galactic center}},
  \href{https://doi.org/10.1103/PhysRevD.82.064009}{\emph{\prd} {\bfseries 82}
  (2010) 064009} [\href{https://arxiv.org/abs/1004.0379}{{\ttfamily
  1004.0379}}].

\bibitem{Eiroa2012PRD86.083009}
E.~F. {Eiroa} and C.~M. {Sendra}, \emph{{Gravitational lensing by massless
  braneworld black holes}},
  \href{https://doi.org/10.1103/PhysRevD.86.083009}{\emph{\prd} {\bfseries 86}
  (2012) 083009} [\href{https://arxiv.org/abs/1207.5502}{{\ttfamily
  1207.5502}}].

\bibitem{Eiroa2013PRD88.103007}
E.~F. {Eiroa} and C.~M. {Sendra}, \emph{{Regular phantom black hole
  gravitational lensing}},
  \href{https://doi.org/10.1103/PhysRevD.88.103007}{\emph{\prd} {\bfseries 88}
  (2013) 103007} [\href{https://arxiv.org/abs/1308.5959}{{\ttfamily
  1308.5959}}].

\bibitem{Zhao2017EPJC77.272}
S.-S. {Zhao} and Y.~{Xie}, \emph{{Strong deflection gravitational lensing by a
  modified Hayward black hole}},
  \href{https://doi.org/10.1140/epjc/s10052-017-4850-5}{\emph{\epjc} {\bfseries
  77} (2017) 272} [\href{https://arxiv.org/abs/1704.02434}{{\ttfamily
  1704.02434}}].

\bibitem{Zhao2017PLB774.357}
S.-S. {Zhao} and Y.~{Xie}, \emph{{Strong deflection lensing by a Lee-Wick black
  hole}}, \href{https://doi.org/10.1016/j.physletb.2017.09.090}{\emph{\plb}
  {\bfseries 774} (2017) 357}.

\bibitem{Bozza2003PRD67.103006}
V.~{Bozza}, \emph{{Quasiequatorial gravitational lensing by spinning black
  holes in the strong field limit}},
  \href{https://doi.org/10.1103/PhysRevD.67.103006}{\emph{\prd} {\bfseries 67}
  (2003) 103006} [\href{https://arxiv.org/abs/gr-qc/0210109}{{\ttfamily
  gr-qc/0210109}}].

\bibitem{Vazquez2004NCimB119.489}
S.~E. {V{\'a}zquez} and E.~P. {Esteban}, \emph{{Strong-field gravitational
  lensing by a Kerr black hole}},
  \href{https://doi.org/10.1393/ncb/i2004-10121-y}{\emph{Nuovo Cimento B Ser.}
  {\bfseries 119} (2004) 489}
  [\href{https://arxiv.org/abs/gr-qc/0308023}{{\ttfamily gr-qc/0308023}}].

\bibitem{Bozza2005PRD72.083003}
V.~{Bozza}, F.~{de Luca}, G.~{Scarpetta} and M.~{Sereno}, \emph{{Analytic Kerr
  black hole lensing for equatorial observers in the strong deflection limit}},
  \href{https://doi.org/10.1103/PhysRevD.72.083003}{\emph{\prd} {\bfseries 72}
  (2005) 083003} [\href{https://arxiv.org/abs/gr-qc/0507137}{{\ttfamily
  gr-qc/0507137}}].

\bibitem{Bozza2006PRD74.063001}
V.~{Bozza}, F.~{de Luca} and G.~{Scarpetta}, \emph{{Kerr black hole lensing for
  generic observers in the strong deflection limit}},
  \href{https://doi.org/10.1103/PhysRevD.74.063001}{\emph{\prd} {\bfseries 74}
  (2006) 063001} [\href{https://arxiv.org/abs/gr-qc/0604093}{{\ttfamily
  gr-qc/0604093}}].

\bibitem{Bozza2007PRD76.083008}
V.~{Bozza} and G.~{Scarpetta}, \emph{{Strong deflection limit of black hole
  gravitational lensing with arbitrary source distances}},
  \href{https://doi.org/10.1103/PhysRevD.76.083008}{\emph{\prd} {\bfseries 76}
  (2007) 083008} [\href{https://arxiv.org/abs/0705.0246}{{\ttfamily
  0705.0246}}].

\bibitem{Bozza2008PRD78.063014}
V.~{Bozza}, \emph{{Optical caustics of Kerr spacetime: The full structure}},
  \href{https://doi.org/10.1103/PhysRevD.78.063014}{\emph{\prd} {\bfseries 78}
  (2008) 063014} [\href{https://arxiv.org/abs/0806.4102}{{\ttfamily
  0806.4102}}].

\bibitem{Chen2010CQG27.225006}
S.~{Chen} and J.~{Jing}, \emph{{Geodetic precession and strong gravitational
  lensing in dynamical Chern-Simons-modified gravity}},
  \href{https://doi.org/10.1088/0264-9381/27/22/225006}{\emph{\cqg} {\bfseries
  27} (2010) 225006} [\href{https://arxiv.org/abs/1005.1325}{{\ttfamily
  1005.1325}}].

\bibitem{Chen2011PRD83.124019}
S.~{Chen}, Y.~{Liu} and J.~{Jing}, \emph{{Strong gravitational lensing in a
  squashed Kaluza-Klein G{\"o}del black hole}},
  \href{https://doi.org/10.1103/PhysRevD.83.124019}{\emph{\prd} {\bfseries 83}
  (2011) 124019} [\href{https://arxiv.org/abs/1102.0086}{{\ttfamily
  1102.0086}}].

\bibitem{Kraniotis2011CQG28.085021}
G.~V. {Kraniotis}, \emph{{Precise analytic treatment of Kerr and Kerr-(anti) de
  Sitter black holes as gravitational lenses}},
  \href{https://doi.org/10.1088/0264-9381/28/8/085021}{\emph{\cqg} {\bfseries
  28} (2011) 085021} [\href{https://arxiv.org/abs/1009.5189}{{\ttfamily
  1009.5189}}].

\bibitem{Chen2012PRD85.124029}
S.~{Chen} and J.~{Jing}, \emph{{Strong gravitational lensing by a rotating
  non-Kerr compact object}},
  \href{https://doi.org/10.1103/PhysRevD.85.124029}{\emph{\prd} {\bfseries 85}
  (2012) 124029} [\href{https://arxiv.org/abs/1204.2468}{{\ttfamily
  1204.2468}}].

\bibitem{Kraniotis2014GRG46.1818}
G.~V. {Kraniotis}, \emph{{Gravitational lensing and frame dragging of light in
  the Kerr-Newman and the Kerr-Newman (anti) de Sitter black hole spacetimes}},
  \href{https://doi.org/10.1007/s10714-014-1818-8}{\emph{\grg} {\bfseries 46}
  (2014) 1818} [\href{https://arxiv.org/abs/1401.7118}{{\ttfamily 1401.7118}}].

\bibitem{Ji2014JHEP03.089}
L.~{Ji}, S.~{Chen} and J.~{Jing}, \emph{{Strong gravitational lensing in a
  rotating Kaluza-Klein black hole with squashed horizons}},
  \href{https://doi.org/10.1007/JHEP03(2014)089}{\emph{\jhep} {\bfseries 3}
  (2014) 89} [\href{https://arxiv.org/abs/1312.4128}{{\ttfamily 1312.4128}}].

\bibitem{Cunha2015PRL115.211102}
P.~V.~P. {Cunha}, C.~A.~R. {Herdeiro}, E.~{Radu} and H.~F. {R{\'u}narsson},
  \emph{{Shadows of Kerr Black Holes with Scalar Hair}},
  \href{https://doi.org/10.1103/PhysRevLett.115.211102}{\emph{\prl} {\bfseries
  115} (2015) 211102} [\href{https://arxiv.org/abs/1509.00021}{{\ttfamily
  1509.00021}}].

\bibitem{Wang2016JCAP11.020}
S.~{Wang}, S.~{Chen} and J.~{Jing}, \emph{{Strong gravitational lensing by a
  Konoplya-Zhidenko rotating non-Kerr compact object}},
  \href{https://doi.org/10.1088/1475-7516/2016/11/020}{\emph{\jcap} {\bfseries
  11} (2016) 020} [\href{https://arxiv.org/abs/1609.00802}{{\ttfamily
  1609.00802}}].

\bibitem{Bin-Nun2010PRD81.123011}
A.~Y. {Bin-Nun}, \emph{{Relativistic images in Randall-Sundrum II braneworld
  lensing}}, \href{https://doi.org/10.1103/PhysRevD.81.123011}{\emph{\prd}
  {\bfseries 81} (2010) 123011}
  [\href{https://arxiv.org/abs/0912.2081}{{\ttfamily 0912.2081}}].

\bibitem{Gyulchev2007PRD75.023006}
G.~N. {Gyulchev} and S.~S. {Yazadjiev}, \emph{{Kerr-Sen dilaton-axion black
  hole lensing in the strong deflection limit}},
  \href{https://doi.org/10.1103/PhysRevD.75.023006}{\emph{\prd} {\bfseries 75}
  (2007) 023006} [\href{https://arxiv.org/abs/gr-qc/0611110}{{\ttfamily
  gr-qc/0611110}}].

\bibitem{Gyulchev2013PRD87.063005}
G.~N. {Gyulchev} and I.~Z. {Stefanov}, \emph{{Gravitational lensing by phantom
  black holes}}, \href{https://doi.org/10.1103/PhysRevD.87.063005}{\emph{\prd}
  {\bfseries 87} (2013) 063005}
  [\href{https://arxiv.org/abs/1211.3458}{{\ttfamily 1211.3458}}].

\bibitem{Zhao2016JCAP07.007}
S.-S. {Zhao} and Y.~{Xie}, \emph{{Strong field gravitational lensing by a
  charged Galileon black hole}},
  \href{https://doi.org/10.1088/1475-7516/2016/07/007}{\emph{\jcap} {\bfseries
  07} (2016) 007} [\href{https://arxiv.org/abs/1603.00637}{{\ttfamily
  1603.00637}}].

\bibitem{Cavalcanti2016CQG33.215007}
R.~T. {Cavalcanti}, A.~{Goncalves da Silva} and R.~{da Rocha}, \emph{{Strong
  deflection limit lensing effects in the minimal geometric deformation and
  Casadio-Fabbri-Mazzacurati solutions}},
  \href{https://doi.org/10.1088/0264-9381/33/21/215007}{\emph{\cqg} {\bfseries
  33} (2016) 215007} [\href{https://arxiv.org/abs/1605.01271}{{\ttfamily
  1605.01271}}].

\bibitem{Chakraborty2017JCAP07.045}
S.~{Chakraborty} and S.~{SenGupta}, \emph{{Strong gravitational lensing---a
  probe for extra dimensions and Kalb-Ramond field}},
  \href{https://doi.org/10.1088/1475-7516/2017/07/045}{\emph{\jcap} {\bfseries
  7} (2017) 045} [\href{https://arxiv.org/abs/1611.06936}{{\ttfamily
  1611.06936}}].

\bibitem{Gyulchev2008PRD78.083004}
G.~N. {Gyulchev} and S.~S. {Yazadjiev}, \emph{{Gravitational lensing by
  rotating naked singularities}},
  \href{https://doi.org/10.1103/PhysRevD.78.083004}{\emph{\prd} {\bfseries 78}
  (2008) 083004} [\href{https://arxiv.org/abs/0806.3289}{{\ttfamily
  0806.3289}}].

\bibitem{Sahu2012PRD86.063010}
S.~{Sahu}, M.~{Patil}, D.~{Narasimha} and P.~S. {Joshi}, \emph{{Can strong
  gravitational lensing distinguish naked singularities from black holes?}},
  \href{https://doi.org/10.1103/PhysRevD.86.063010}{\emph{\prd} {\bfseries 86}
  (2012) 063010} [\href{https://arxiv.org/abs/1206.3077}{{\ttfamily
  1206.3077}}].

\bibitem{Sahu2013PRD88.103002}
S.~{Sahu}, M.~{Patil}, D.~{Narasimha} and P.~S. {Joshi}, \emph{{Time delay
  between relativistic images as a probe of cosmic censorship}},
  \href{https://doi.org/10.1103/PhysRevD.88.103002}{\emph{\prd} {\bfseries 88}
  (2013) 103002} [\href{https://arxiv.org/abs/1310.5350}{{\ttfamily
  1310.5350}}].

\bibitem{Virbhadra2002PRD65.103004}
K.~S. {Virbhadra} and G.~F. {Ellis}, \emph{{Gravitational lensing by naked
  singularities}},
  \href{https://doi.org/10.1103/PhysRevD.65.103004}{\emph{\prd} {\bfseries 65}
  (2002) 103004}.

\bibitem{Kuhfittig2014EPJC74.2818}
P.~K.~F. {Kuhfittig}, \emph{{Gravitational lensing of wormholes in the galactic
  halo region}},
  \href{https://doi.org/10.1140/epjc/s10052-014-2818-2}{\emph{\epjc} {\bfseries
  74} (2014) 2818} [\href{https://arxiv.org/abs/1311.2274}{{\ttfamily
  1311.2274}}].

\bibitem{Kuhfittig2015arXiv1501.06085}
P.~K.~F. {Kuhfittig}, \emph{{Gravitational lensing of wormholes in
  noncommutative geometry}}, {\emph{arXiv:1501.06085} (2015) }
  [\href{https://arxiv.org/abs/1501.06085}{{\ttfamily 1501.06085}}].

\bibitem{Nandi2006PRD74.024020}
K.~K. {Nandi}, Y.-Z. {Zhang} and A.~V. {Zakharov}, \emph{{Gravitational lensing
  by wormholes}}, \href{https://doi.org/10.1103/PhysRevD.74.024020}{\emph{\prd}
  {\bfseries 74} (2006) 024020}
  [\href{https://arxiv.org/abs/gr-qc/0602062}{{\ttfamily gr-qc/0602062}}].

\bibitem{Tsukamoto2012PRD86.104062}
N.~{Tsukamoto}, T.~{Harada} and K.~{Yajima}, \emph{{Can we distinguish between
  black holes and wormholes by their Einstein-ring systems?}},
  \href{https://doi.org/10.1103/PhysRevD.86.104062}{\emph{\prd} {\bfseries 86}
  (2012) 104062} [\href{https://arxiv.org/abs/1207.0047}{{\ttfamily
  1207.0047}}].

\bibitem{Tsukamoto2016PRD94.124001}
N.~Tsukamoto, \emph{Strong deflection limit analysis and gravitational lensing
  of an ellis wormhole},
  \href{https://doi.org/10.1103/PhysRevD.94.124001}{\emph{\prd} {\bfseries 94}
  (2016) 124001}.

\bibitem{Eiroa2014EPJC74.3171}
E.~F. {Eiroa} and C.~M. {Sendra}, \emph{{Strong deflection lensing by charged
  black holes in scalar-tensor gravity}},
  \href{https://doi.org/10.1140/epjc/s10052-014-3171-1}{\emph{\epjc} {\bfseries
  74} (2014) 3171} [\href{https://arxiv.org/abs/1408.3390}{{\ttfamily
  1408.3390}}].

\bibitem{Sotani2015PRD92.044052}
H.~Sotani and U.~Miyamoto, \emph{Strong gravitational lensing by an
  electrically charged black hole in eddington-inspired born-infeld gravity},
  \href{https://doi.org/10.1103/PhysRevD.92.044052}{\emph{\prd} {\bfseries 92}
  (2015) 044052}.

\bibitem{Chen2015JCAP10.002}
S.~{Chen} and J.~{Jing}, \emph{{Strong gravitational lensing for the photons
  coupled to Weyl tensor in a Schwarzschild black hole spacetime}},
  \href{https://doi.org/10.1088/1475-7516/2015/10/002}{\emph{\jcap} {\bfseries
  10} (2015) 002} [\href{https://arxiv.org/abs/1502.01088}{{\ttfamily
  1502.01088}}].

\bibitem{Lu2016EPJC76.357}
X.~{Lu}, F.-W. {Yang} and Y.~{Xie}, \emph{{Strong gravitational field time
  delay for photons coupled to Weyl tensor in a Schwarzschild black hole}},
  \href{https://doi.org/10.1140/epjc/s10052-016-4218-2}{\emph{\epjc} {\bfseries
  76} (2016) 357} [\href{https://arxiv.org/abs/1606.02932}{{\ttfamily
  1606.02932}}].

\bibitem{Chen2017PRD95.104017}
S.~{Chen}, S.~{Wang}, Y.~{Huang}, J.~{Jing} and S.~{Wang}, \emph{{Strong
  gravitational lensing for the photons coupled to a Weyl tensor in a Kerr
  black hole spacetime}},
  \href{https://doi.org/10.1103/PhysRevD.95.104017}{\emph{\prd} {\bfseries 95}
  (2017) 104017} [\href{https://arxiv.org/abs/1611.08783}{{\ttfamily
  1611.08783}}].

\bibitem{Zhang2018EPJC78.796}
R.~{Zhang} and J.~{Jing}, \emph{{Strong gravitational lensing for photons
  coupled to Weyl tensor in a regular phantom black hole}},
  \href{https://doi.org/10.1140/epjc/s10052-018-6272-4}{\emph{\epjc} {\bfseries
  78} (2018) 796}.

\bibitem{Chen2018EPJC78.981}
S.~{Chen}, L.~{Zhang} and J.~{Jing}, \emph{{A new asymptotical flat and
  spherically symmetric solution in the generalized
  Einstein-Cartan-Kibble-Sciama gravity and gravitational lensing}},
  \href{https://doi.org/10.1140/epjc/s10052-018-6466-9}{\emph{\epjc} {\bfseries
  78} (2018) 981} [\href{https://arxiv.org/abs/1804.05004}{{\ttfamily
  1804.05004}}].

\bibitem{Badia2017EPJC77.779}
J.~{Bad{\'{\i}}a} and E.~F. {Eiroa}, \emph{{Gravitational lensing by a
  Horndeski black hole}},
  \href{https://doi.org/10.1140/epjc/s10052-017-5376-6}{\emph{\epjc} {\bfseries
  77} (2017) 779} [\href{https://arxiv.org/abs/1707.02970}{{\ttfamily
  1707.02970}}].

\bibitem{Petters2003MNRAS338.457}
A.~O. {Petters}, \emph{{On relativistic corrections to microlensing effects:
  applications to the Galactic black hole}},
  \href{https://doi.org/10.1046/j.1365-8711.2003.06065.x}{\emph{\mnras}
  {\bfseries 338} (2003) 457}
  [\href{https://arxiv.org/abs/astro-ph/0208500}{{\ttfamily
  astro-ph/0208500}}].

\bibitem{Horvath2011PRD84.083006}
Z.~{Horv{\'a}th}, L.~{\'A}. {Gergely}, Z.~{Keresztes}, T.~{Harko} and F.~S.~N.
  {Lobo}, \emph{{Constraining Ho{\v r}ava-Lifshitz gravity by weak and strong
  gravitational lensing}},
  \href{https://doi.org/10.1103/PhysRevD.84.083006}{\emph{\prd} {\bfseries 84}
  (2011) 083006} [\href{https://arxiv.org/abs/1105.0765}{{\ttfamily
  1105.0765}}].

\bibitem{Ray2003PRD68.084004}
S.~{Ray}, A.~L. {Esp{\'{\i}}ndola}, M.~{Malheiro}, J.~P. {Lemos} and V.~T.
  {Zanchin}, \emph{{Electrically charged compact stars and formation of charged
  black holes}}, \href{https://doi.org/10.1103/PhysRevD.68.084004}{\emph{\prd}
  {\bfseries 68} (2003) 084004}
  [\href{https://arxiv.org/abs/astro-ph/0307262}{{\ttfamily
  astro-ph/0307262}}].

\bibitem{Wald1974PRD10.1680}
R.~M. {Wald}, \emph{{Black hole in a uniform magnetic field}},
  \href{https://doi.org/10.1103/PhysRevD.10.1680}{\emph{\prd} {\bfseries 10}
  (1974) 1680}.

\bibitem{Zajacek2018MNRAS480.4408}
M.~{Zaja{\v c}ek}, A.~{Tursunov}, A.~{Eckart} and S.~{Britzen}, \emph{{On the
  charge of the Galactic centre black hole}},
  \href{https://doi.org/10.1093/mnras/sty2182}{\emph{\mnras} {\bfseries 480}
  (2018) 4408} [\href{https://arxiv.org/abs/1808.07327}{{\ttfamily
  1808.07327}}].

\bibitem{Reissner1916AnP355.106}
H.~{Reissner}, \emph{{{\"U}ber die Eigengravitation des elektrischen Feldes
  nach der Einsteinschen Theorie}},
  \href{https://doi.org/10.1002/andp.19163550905}{\emph{Annalen der Physik}
  {\bfseries 355} (1916) 106}.

\bibitem{Nordstrom1918KNAB20.1238}
G.~{Nordstr{\"o}m}, \emph{{On the Energy of the Gravitation field in Einstein's
  Theory}}, {\emph{Koninklijke Nederlandse Akademie van Wetenschappen}
  {\bfseries 20} (1918) 1238}.

\bibitem{Maartens2004LRR7.7}
R.~{Maartens}, \emph{{Brane-World Gravity}}, {\emph{Living Reviews in
  Relativity} {\bfseries 7} (2004) 7}
  [\href{https://arxiv.org/abs/gr-qc/0312059}{{\ttfamily gr-qc/0312059}}].

\bibitem{Dadhich2000PLB487.1}
N.~{Dadhich}, R.~{Maartens}, P.~{Papadopoulos} and V.~{Rezania}, \emph{{Black
  holes on the brane}},
  \href{https://doi.org/10.1016/S0370-2693(00)00798-X}{\emph{Physics Letters B}
  {\bfseries 487} (2000) 1}
  [\href{https://arxiv.org/abs/hep-th/0003061}{{\ttfamily hep-th/0003061}}].

\bibitem{GRAVITYCol2017A&A602.A94}
{GRAVITY Collaboration}, \emph{{First light for GRAVITY: Phase referencing
  optical interferometry for the Very Large Telescope Interferometer}},
  \href{https://doi.org/10.1051/0004-6361/201730838}{\emph{\aap} {\bfseries
  602} (2017) A94} [\href{https://arxiv.org/abs/1705.02345}{{\ttfamily
  1705.02345}}].

\bibitem{Doeleman2008Nature455.78}
S.~S. {Doeleman}, J.~{Weintroub}, A.~E.~E. {Rogers}, R.~{Plambeck},
  R.~{Freund}, R.~P.~J. {Tilanus} et~al., \emph{{Event-horizon-scale structure
  in the supermassive black hole candidate at the Galactic Centre}},
  \href{https://doi.org/10.1038/nature07245}{\emph{\nat} {\bfseries 455} (2008)
  78} [\href{https://arxiv.org/abs/0809.2442}{{\ttfamily 0809.2442}}].

\bibitem{Weinberg1972Book}
S.~{Weinberg}, \emph{{Gravitation and Cosmology: Principles and Applications of
  the General Theory of Relativity}}. {Wiley}, {New York}, July, 1972.

\bibitem{Refsdal1964MNRAS128.295}
S.~{Refsdal}, \emph{{The gravitational lens effect}},
  \href{https://doi.org/10.1093/mnras/128.4.295}{\emph{\mnras} {\bfseries 128}
  (1964) 295}.

\bibitem{Pang2019CQG36.065012}
X.~{Pang} and J.~{Jia}, \emph{{Gravitational lensing of massive particles in
  Reissner-Nordstr\"om black hole spacetime}},
  \href{https://doi.org/10.1088/1361-6382/ab0512}{\emph{\cqg} {\bfseries 36}
  (2019) {065012}}.

\bibitem{Gillessen2017ApJ837.30}
S.~{Gillessen}, P.~M. {Plewa}, F.~{Eisenhauer}, R.~{Sari} and et~al., \emph{{An
  Update on Monitoring Stellar Orbits in the Galactic Center}},
  \href{https://doi.org/10.3847/1538-4357/aa5c41}{\emph{\apj} {\bfseries 837}
  (2017) 30}.

\bibitem{Huang2018ApJ868.L39}
C.~X. {Huang}, J.~{Burt}, A.~{Vanderburg}, M.~N. {G{\"u}nther}, A.~{Shporer},
  J.~A. {Dittmann} et~al., \emph{{TESS Discovery of a Transiting Super-Earth in
  the pi Mensae System}},
  \href{https://doi.org/10.3847/2041-8213/aaef91}{\emph{\apjl} {\bfseries 868}
  (2018) L39} [\href{https://arxiv.org/abs/1809.05967}{{\ttfamily
  1809.05967}}].

\bibitem{Claudel2001JMP42.818}
C.-M. {Claudel}, K.~S. {Virbhadra} and G.~F.~R. {Ellis}, \emph{{The geometry of
  photon surfaces}}, \href{https://doi.org/10.1063/1.1308507}{\emph{\jmp}
  {\bfseries 42} (2001) 818}
  [\href{https://arxiv.org/abs/gr-qc/0005050}{{\ttfamily gr-qc/0005050}}].

\bibitem{Zakharov2005AA442.795}
A.~F. {Zakharov}, F.~{de Paolis}, G.~{Ingrosso} and A.~A. {Nucita},
  \emph{{Direct measurements of black hole charge with future astrometrical
  missions}}, \href{https://doi.org/10.1051/0004-6361:20053432}{\emph{\aap}
  {\bfseries 442} (2005) 795}.

\bibitem{Horvath2013AN334.1047}
Z.~{Horv{\'a}th} and L.~{\'A}. {Gergely}, \emph{{Black hole tidal charge
  constrained by strong gravitational lensing}},
  \href{https://doi.org/10.1002/asna.201211992}{\emph{\an} {\bfseries 334}
  (2013) 1047} [\href{https://arxiv.org/abs/1203.6576}{{\ttfamily 1203.6576}}].

\bibitem{Zakharov2014PRD90.062007}
A.~F. {Zakharov}, \emph{{Constraints on a charge in the Reissner-Nordstr{\"o}m
  metric for the black hole at the Galactic Center}},
  \href{https://doi.org/10.1103/PhysRevD.90.062007}{\emph{\prd} {\bfseries 90}
  (2014) 062007}.

\bibitem{Ibanez1983A&A124.175}
J.~{Ibanez}, \emph{{Gravitational lenses with angular momentum}}, {\emph{\aap}
  {\bfseries 124} (1983) 175}.

\bibitem{Bray1986PRD34.367}
I.~{Bray}, \emph{{Kerr black hole as a gravitational lens}},
  \href{https://doi.org/10.1103/PhysRevD.34.367}{\emph{\prd} {\bfseries 34}
  (1986) 367}.

\bibitem{Klioner1991SvA35.523}
S.~A. {Klioner}, \emph{{Influence of the Quadrupole Field and Rotation of
  Objects on Light Propagation}}, {\emph{\sovast} {\bfseries 35} (1991) 523}.

\bibitem{Glicenstein1999A&A343.1025}
J.~F. {Glicenstein}, \emph{{Gravitational lensing by rotating stars}},
  {\emph{\aap} {\bfseries 343} (1999) 1025}.

\bibitem{Sereno2006PRD74.123009}
M.~{Sereno} and F.~{de Luca}, \emph{{Analytical Kerr black hole lensing in the
  weak deflection limit}},
  \href{https://doi.org/10.1103/PhysRevD.74.123009}{\emph{\prd} {\bfseries 74}
  (2006) 123009} [\href{https://arxiv.org/abs/astro-ph/0609435}{{\ttfamily
  astro-ph/0609435}}].

\bibitem{Werner2007PRD76.064024}
M.~C. {Werner} and A.~O. {Petters}, \emph{{Magnification relations for Kerr
  lensing and testing cosmic censorship}},
  \href{https://doi.org/10.1103/PhysRevD.76.064024}{\emph{\prd} {\bfseries 76}
  (2007) 064024} [\href{https://arxiv.org/abs/0706.0132}{{\ttfamily
  0706.0132}}].

\bibitem{Sereno2008PRD78.023008}
M.~{Sereno} and F.~{de Luca}, \emph{{Primary caustics and critical points
  behind a Kerr black hole}},
  \href{https://doi.org/10.1103/PhysRevD.78.023008}{\emph{\prd} {\bfseries 78}
  (2008) 023008} [\href{https://arxiv.org/abs/0710.5923}{{\ttfamily
  0710.5923}}].

\bibitem{Aazami2011JMP52.092502}
A.~B. {Aazami}, C.~R. {Keeton} and A.~O. {Petters}, \emph{{Lensing by Kerr
  black holes. I. General lens equation and magnification formula}},
  \href{https://doi.org/10.1063/1.3642614}{\emph{\jmp} {\bfseries 52} (2011)
  092502} [\href{https://arxiv.org/abs/1102.4300}{{\ttfamily 1102.4300}}].

\bibitem{Aazami2011JMP52.102501}
A.~B. {Aazami}, C.~R. {Keeton} and A.~O. {Petters}, \emph{{Lensing by Kerr
  black holes. II: Analytical study of quasi-equatorial lensing observables}},
  \href{https://doi.org/10.1063/1.3642616}{\emph{\jmp} {\bfseries 52} (2011)
  102501} [\href{https://arxiv.org/abs/1102.4304}{{\ttfamily 1102.4304}}].

\bibitem{He2014IJMPD23.1450031}
G.~{He} and W.~{Lin}, \emph{{Roto-Translational Effects on Deflection of Light
  and Particle by Moving Kerr Black Hole}},
  \href{https://doi.org/10.1142/S021827181450031X}{\emph{\ijmpd} {\bfseries 23}
  (2014) 1450031}.

\bibitem{He2014IJMPD23.1450079}
G.~{He}, C.~{Jiang} and W.~{Lin}, \emph{{Second post-Minkowskian metric for a
  moving Kerr black hole}},
  \href{https://doi.org/10.1142/S0218271814500795}{\emph{\ijmpd} {\bfseries 23}
  (2014) 1450079}.

\bibitem{Deng2015IJMPD24.1550056}
X.-M. {Deng}, \emph{{The second post-Newtonian light propagation and its
  astrometric measurement in the solar system}},
  \href{https://doi.org/10.1142/S021827181550056X}{\emph{\ijmpd} {\bfseries 24}
  (2015) 1550056} [\href{https://arxiv.org/abs/1504.04084}{{\ttfamily
  1504.04084}}].

\bibitem{He2015RAA15.646}
G.-S. {He} and W.-B. {Lin}, \emph{{Second post-Minkowskian order harmonic
  metric for a moving Kerr-Newman black hole}},
  \href{https://doi.org/10.1088/1674-4527/15/5/003}{\emph{\raa} {\bfseries 15}
  (2015) 646}.

\bibitem{Deng2016IJMPD25.1650082}
X.-M. {Deng}, \emph{{The second post-Newtonian light propagation and its
  astrometric measurement in the Solar System: Light time and frequency
  shift}}, \href{https://doi.org/10.1142/S0218271816500826}{\emph{\ijmpd}
  {\bfseries 25} (2016) 1650082}.

\bibitem{He2016PRD93.023005}
G.~{He} and W.~{Lin}, \emph{{Second order Kerr-Newman time delay}},
  \href{https://doi.org/10.1103/PhysRevD.93.023005}{\emph{\prd} {\bfseries 93}
  (2016) 023005}.

\bibitem{He2016PRD94.063011}
G.~{He} and W.~{Lin}, \emph{{Second-order time delay by a radially moving
  Kerr-Newman black hole}},
  \href{https://doi.org/10.1103/PhysRevD.94.063011}{\emph{\prd} {\bfseries 94}
  (2016) 063011}.

\bibitem{He2017CQG34.105006}
G.~{He} and W.~{Lin}, \emph{{Analytical derivation of second-order deflection
  in the equatorial plane of a radially moving Kerr-Newman black hole}},
  \href{https://doi.org/10.1088/1361-6382/aa691d}{\emph{\cqg} {\bfseries 34}
  (2017) 105006}.

\bibitem{Bardeen1973BH.215}
J.~M. {Bardeen}, \emph{{Timelike and null geodesics in the Kerr metric.}},  in
  \emph{Black Holes (Les Astres Occlus)}, C.~{Dewitt} and B.~S. {Dewitt}, eds.,
  pp.~215--239, {Gordon and Breach}, 1973.

\bibitem{Iyer2009PRD80.124023}
S.~V. {Iyer} and E.~C. {Hansen}, \emph{{Light's bending angle in the equatorial
  plane of a Kerr black hole}},
  \href{https://doi.org/10.1103/PhysRevD.80.124023}{\emph{\prd} {\bfseries 80}
  (2009) 124023} [\href{https://arxiv.org/abs/0907.5352}{{\ttfamily
  0907.5352}}].

\end{thebibliography}\endgroup



\end{document}